\documentclass[acmsmall, screen]{acmart}

\usepackage{amsmath,amssymb,amsfonts}
\usepackage{hyperref}
\hypersetup{colorlinks=true}
\def\equationautorefname~#1\null{%
  Equation~(#1)\null
}
\usepackage{algorithm}
\usepackage[noend]{algpseudocode}
\usepackage{colortbl}
\usepackage{xcolor}
\usepackage{colortbl}%

\usepackage{capt-of,lipsum}
\definecolor{mycolor}{rgb}{0.95, 0.985, 0.93}
\doublerulesepcolor{mycolor}
\usepackage{xspace}
\usepackage{array}
\usepackage{color}
\usepackage{tcolorbox}
\definecolor{mGray1}{rgb}{0.9,0.9,0.9}
\definecolor{mGray}{rgb}{0.5,0.5,0.5}
\definecolor{commentcolor}{rgb}{0.6,0.6,0.6}
\usepackage{enumitem}
\newtheorem{definition}{Definition}
\usepackage{pifont}
\usepackage{xcolor}
\usepackage{listings}
\usepackage{subcaption}
\usepackage{booktabs}
\usepackage{stmaryrd}
\usepackage{graphicx}
\usepackage{wrapfig}
\newcommand{\m}{\mathit}

\newcommand{\relation}{R} 
\newcommand{\Obj}{o} 
\newcommand{\Subj}{s} 
\newcommand{\OBJ}{O} 
\newcommand{\SUBJ}{S} 
\newcommand{\Prolog}{\mathcal{P}}
\newcommand{\drule}{Q}

\newcommand{\hornarrow}{\,\text{:--}\,}

\newcommand{\deriveRules}[3]{#1\,{\hookrightarrow}\,(#2, #3)}
\newcommand{\nm}{\m{nm}}
\newcommand{\entity}{\m{entity}}
\newcommand{\groundTruthTriples}{\widetilde{\relation}_{\m{ground}}}
\newcommand{\derivedFacts}{\widetilde{\relation}_{\m{derived}}}

\newcommand{\llmResponse}{\m{Resp}}

\newcommand{\semantic}{\widetilde{SS}}
\newcommand{\similarity}{\m{J\_Sim}}
\newcommand{\rall}{\widetilde{\relation}_{\m{all}}}

\usepackage{makecell}

\newcommand{\commentstyle}[1]{\textcolor{mGray}{\footnotesize{#1}}}

\newcommand\algoref[1]{Algorithm~\textcolor{blue}{\ref{#1}}}

\newcommand{\entityCat}{$\m{EC}$}
\newcommand{\relationCat}{$\m{RC}$}

\algrenewcommand\algorithmicindent{1.2em}

\definecolor{keywordcolor}{rgb}{0.13,0.29,0.53}
\definecolor{stringcolor}{rgb}{0.31,0.60,0.02}
\definecolor{commentcolor}{rgb}{0.56,0.35,0.01}
\definecolor{backcolour}{rgb}{0.95,0.95,0.92}
\newcommand{\head}[1]{{\noindent\textbf{#1}}}

\newcommand{\tool}{\textsc{Drowzee}\xspace}
\newcommand{\instruction}{\textsc{Instruction}\xspace}
\newcommand{\query}{\textsc{Query}\xspace}

\AtBeginDocument{%
  \providecommand\BibTeX{{%
    \normalfont B\kern-0.5em{\scshape i\kern-0.25em b}\kern-0.8em\TeX}}}

\setcopyright{rightsretained}
\acmDOI{10.1145/3689776}
\acmYear{2024}
\acmJournal{PACMPL}
\acmVolume{8}
\acmNumber{OOPSLA2}
\acmArticle{336}
\acmMonth{10}
\acmSubmissionID{oopslab24main-p613-p}
\received{2024-04-06}
\received[accepted]{2024-08-18}

\begin{document}

\title{Drowzee: Metamorphic Testing for Fact-Conflicting Hallucination Detection in Large Language Models}

\author{Ningke Li}
\authornote{Ningke Li and Yuekang Li are co-first authors.}
\affiliation{%
  \institution{Huazhong University of Science and Technology}
  \country{China}
}
\email{lnk\_01@hust.edu.cn}

\author{Yuekang Li}
\authornotemark[1]
\affiliation{%
 \institution{The University of New South Wales}
 \country{Australia}
 }
\email{yuekang.li@unsw.edu.au}

\author{Yi Liu}
\affiliation{%
 \institution{Nanyang Technological University}
 \country{Singapore}
 }
\email{yi009@e.ntu.edu.sg}

\author{Ling Shi}
\affiliation{%
 \institution{Nanyang Technological University}
 \country{Singapore}
 }
\email{ling.shi@ntu.edu.sg}

\author{Kailong Wang}
\authornote{Kailong Wang is the corresponding author.}
\affiliation{%
 \institution{Huazhong University of Science and Technology}
 \country{China}
 }
\email{wangkl@hust.edu.cn}

\author{Haoyu Wang}
\affiliation{%
  \institution{Huazhong University of Science and Technology}
  \country{China}}
\email{haoyuwang@hust.edu.cn}


\begin{abstract}
Large language models (LLMs) have revolutionized language processing, but face critical challenges with security, privacy, and generating hallucinations --- coherent but factually inaccurate outputs. A major issue is fact-conflicting hallucination (FCH), where LLMs produce content contradicting ground truth facts. Addressing FCH is difficult due to two key challenges: \textbf{1)} Automatically constructing and updating benchmark datasets is hard, as existing methods rely on manually curated static benchmarks that cannot cover the broad, evolving spectrum of FCH cases. \textbf{2)} Validating the reasoning behind LLM outputs is inherently difficult, especially for complex logical relations.

To tackle these challenges, we introduce a novel logic-programming-aided metamorphic testing technique for FCH detection. We develop an extensive and extensible framework that constructs a comprehensive factual knowledge base by crawling sources like Wikipedia, seamlessly integrated into \tool{}\footnote{\tool{} is named after a Pokémon~\cite{pokemon} character that nourishes itself by eating dreams. This name symbolizes our tool's capability to detect and potentially further assist in eliminating the hallucinations in LLMs.}. Using logical reasoning rules, we transform and augment this knowledge into a large set of test cases with ground truth answers. We test LLMs on these cases through template-based prompts, requiring them to provide reasoned answers. To validate their reasoning, we propose two semantic-aware oracles that assess the similarity between the semantic structures of the LLM answers and ground truth.
Our approach automatically generates useful test cases and identifies hallucinations across six LLMs within nine domains, with hallucination rates ranging from 24.7\% to 59.8\%. Key findings include LLMs struggling with temporal concepts, out-of-distribution knowledge, and lack of logical reasoning capabilities. The results show that logic-based test cases generated by \tool effectively trigger and detect hallucinations.
To further mitigate the identified FCHs, we explored model editing techniques, which proved effective on a small scale (with edits to fewer than 1000 knowledge pieces). Our findings emphasize the need for continued community efforts to detect and mitigate model hallucinations.

\end{abstract}





\maketitle

\section{Introduction}
\begin{figure}[!ht]
    \centering
    \includegraphics[width=\linewidth]{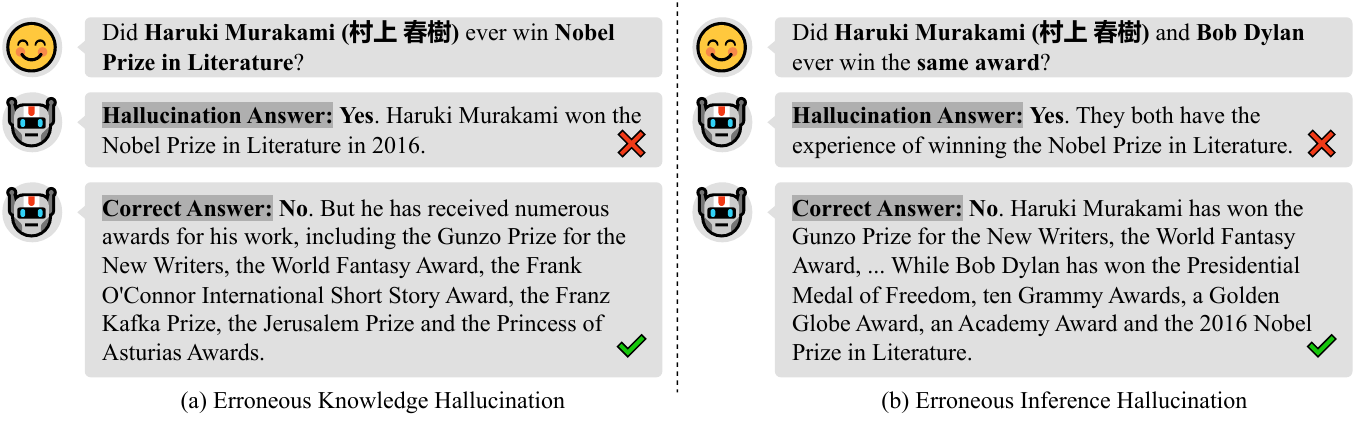}\\
    \caption{A Hallucination Output Example. }
    \label{fig:example1}
\end{figure}

Large Language Models~(LLMs) have brought transformative advancements to the fields of language processing and beyond, showcasing exceptional abilities in text generation and comprehension with wide-ranging applications. However, despite their increasing prevalence, LLMs face critical challenges in security and privacy aspects~\cite{siddiq2023generate, hou2023large, kaddour2023challenges, distillseq, zhang2024glitchproberadvancingeffectivedetection, xu2024largelanguagemodelscyber}, 
heavily impacting their effectiveness and reliability. A particularly notable issue among these is the phenomenon of ``hallucination'', where LLMs produce coherent but factually inaccurate or irrelevant outputs during tasks like problem-solving. This tendency to generate misleading information not only jeopardizes the safety of LLM applications but also raises serious usability concerns. Hallucinations in LLMs take several forms, with ``Fact-conflicting hallucination''~(FCH) being a major concern and the primary type of concern in this paper. FCH is manifested by LLMs generating content that directly contradicts established facts, as exemplified in Figure~\ref{fig:example1}. When an LLM incorrectly believes ``Haruki Murakami won the
Nobel Prize in Literature in 2016'', deviating from the correct answer of ``Haruki Murakami has not won the Nobel Prize but other numerous
awards for his work in Literature''. Such misinformation dissemination leads to significant user confusion, eroding the trust and reliability that are crucial in various LLM applications.

To address the issue of hallucinations in LLMs, recent studies have introduced a range of methods for their detection and testing. A common and straightforward approach involves creating extensive benchmarks tailored for this purpose. Datasets such as TruthfulQA~\cite{lin-etal-2022-truthfulqa}, HaluEval~\cite{HaluEval}, and KoLA~\cite{yu2023kola} have been designed to evaluate hallucinations across different contexts, including question-answering, summarization, and knowledge graphs. 
Despite the value of these manually labeled datasets, the current techniques for hallucination detection and testing heavily rely on naive and semi-automatic methods, such as string matching, manual validation, or utilizing another LLM for confirmation.
This current research landscape in LLM, however, presents a critical gap in automatically and effectively testing FCHs.
The main obstacle in testing for FCH is the absence of dedicated ground truth datasets and specific testing frameworks. Unlike other types of hallucinations~(e.g., input-conflicting and context-conflicting hallucinations, to be detailed in Section~\ref{subsec:cat}) which can be identified through checks for semantic consistency, FCH demands the verification of the content's factual accuracy against external sources of knowledge or databases. This requirement makes the process particularly challenging and resource-intensive, especially for tasks processing contents with inherent logical connections.


Bridging the identified research gap in the literature necessitates an exploration of the inherent challenges faced in detecting FCHs, which are crucial for advancing and enhancing the reliability of LLMs.
\textbf{\textit{Challenge\#1: difficulty in automatically constructing and updating benchmark datasets.}} Predominantly, existing methodologies are anchored to manually curated benchmarks. While these benchmarks are effective in detecting certain types of hallucinations, they fall short in encompassing the broad and dynamic spectrum of fact-conflicting scenarios inherent to LLMs.
Meanwhile, the need for frequent updates to benchmark data, due to the ever-evolving nature of knowledge, imposes a significant and continuous maintenance effort. The reliance on benchmark datasets thus restricts the detection techniques' adaptability, scalability, and worse, detection capability.
\textbf{\textit{Challenge\#2: difficulty in automatically validating answers from LLM outputs.}} Even when LLMs produce correct final answers, the outputs may not represent the true reasoning process behind them, potentially masking false understanding – a source of FCH hallucination. Automatically validating the reasoning process, especially those involving complex logic relations, is inherently difficult. Furthermore, the consistency in the quality of benchmark questions can vary due to the differing levels of experience and skill among human experts creating them, introducing noise, particularly in data labeling and result validation stages.
\head{Our Work.}
To address limitations in the existing techniques, we are the first, to the best of our knowledge, to introduce a novel automatic logic-programming-aided metamorphic testing technique for hallucination detection in this work. We have developed an extensive and extensible FCH testing framework, which is based on factual knowledge reasoning and metamorphic testing, seamlessly integrated into \tool.

\tool begins by establishing a comprehensive factual knowledge base, sourced through extensive crawling of information from accessible knowledge bases such as Wikipedia. Each piece of this knowledge acts as a ``seed'' for subsequent transformations. Leveraging logic reasoning relations, 
we transform and augment these seeds, thereby expanding the factual knowledge into a well-established set of question-answer pairs. 
Using the questions and answers in the knowledge set as test cases and ground truth respectively, we construct a reliable and robust FCH testing benchmark. 
This is implemented through a series of well-formulated template-based prompts to test FCH in LLMs. Specifically, we instruct the LLMs to generate their answers to the test cases. To facilitate a thorough evaluation of the reasoning logic behind their responses, we require the LLMs to provide detailed justifications for their answers. For effective and dependable identification of FCH, we introduce two semantic-aware and similarity-based metamorphic oracles. These oracles operate by extracting essential semantic elements from each sentence and mapping out their logical relationships. By assessing the similarity between the constructed logical and semantic structures of the LLM's answers and the ground truth, we can detect FCH by pinpointing answers that significantly diverge from the ground truth.



\head{Results and Findings.}
In evaluating our proposed FCH testing framework and \tool, we undertake comprehensive experiments to evaluate their effectiveness in a wide array of contexts. On the one hand, our evaluation strategy involves deploying \tool across a broad spectrum of topics, sourced from an extensive and diverse range of Wikipedia articles. On the other hand, we examine our framework on a variety of open-source and commercial LLMs, providing a thorough examination of its applicability and performance across different model architectures. 

Our key findings indicate that \tool succeeds in automatically generating useful test cases and identifying hallucination issues of six LLMs across nine domains. 
Using these test sets, we find that hallucination responses generated by different LLMs can vary from 24.7\% to 59.8\%. 
We then categorize these hallucination responses into four types. Through an in-depth analysis, we unveil that the lack of logical reasoning capabilities contributes the most to the FCH issues in LLMs. 
Additionally, we observe that LLMs are particularly prone to generating hallucinations in test cases involving temporal concepts and out-of-distribution knowledge.
Furthermore, we confirm that test cases generated using our logical reasoning rules can effectively trigger and detect hallucination issues in LLMs.
As mitigation, we investigate the use of model editing techniques~\cite{meng2022locating,fastedit} to rectify the identified FCHs. These techniques have shown promising results when applied on a small scale (involving edits up to less than 1000 pieces of knowledge). Our results highlight the importance of ongoing efforts within the community to detect and address issues of hallucination in LLMs.

\head{Contributions.}
We summarize the main contributions of this paper below:
\begin{itemize}
\item \textbf{Development of a novel FCH Testing Framework.} To the best of our knowledge, we are the first to develop a novel testing framework based on logic programming and metamorphic testing to automatically detect FCH issues in LLMs. 
\item \textbf{Construction and Release of Extensive Factual Knowledge Base and Benchmark.} Our work constructs a large-scale benchmark dataset~\cite{drowzee} to facilitate collaborative efforts and future advancements in the detection of FCH.
\item \textbf{Designing and Implementing Innovative Logic-reasoning-based Method for Data Mutation.} We propose and implement five unique logic reasoning rules to mutate and augment the initial seeds from our knowledge base, increasing the diversity and effectiveness of our test scenario.
\item \textbf{Deployment of FCH-specific semantic-aware testing oracles for automatic LLM answer validation.} We propose and implement two automated verification mechanisms~(oracles) that analyze the semantic structure similarity between sentences. These oracles are designed to validate the reasoning logic behind the answers generated by LLMs, hereby reliably detecting the occurrence of FCHs. 

\end{itemize}

\section{Background}\label{sec:background}
\subsection{Hallucination Categorization}\label{subsec:cat}
Hallucination in LLMs can be categorized into three main categories~\cite{yao2023survey,huang2023survey,zhang2023hallucination}, as detailed below. 

\head{Input-Conflicting Hallucination}: This type arises when LLMs produce outputs that are inconsistent with the user’s input. This inconsistency can occur in two ways: either the model's response contradicts the task instructions (reflecting a misunderstanding of user intents) or the generated content contradicts the task input (similar to conventional issues in machine translation and summarization). An example of this would be an LLM replacing a key name or detail in a summary, deviating from the actual content provided by the user.

\head{Context-Conflicting Hallucination}:  In this case, LLMs exhibit contradictions or inconsistencies in lengthy or multi-turn responses. This happens when models lose track of the context or fail to maintain consistency throughout the conversation. Limitations in maintaining long-term memory or identifying relevant context are often the culprits. An instance of context-conflicting hallucination could involve LLMs switching references between two different individuals in a conversation about a specific topic.

\head{Fact-Conflicting Hallucination:} This type of hallucination is the key focus of this paper. It occurs when LLMs generate information that is in direct conflict with established world knowledge. This can be due to various factors introduced at different stages of the LLM lifecycle. For example, as shown in Figure~\ref{fig:example1}, an LLM might provide incorrect historical information in response to a user's query, misleading users who are less knowledgeable about the subject.

In this paper, our primary focus is on fact-conflicting hallucinations, a type of error that carries the potential for more serious consequences by misleading users.

\subsection{Logic Programming}

\begin{figure}[!ht]
    \centering
    \includegraphics[width=\linewidth]{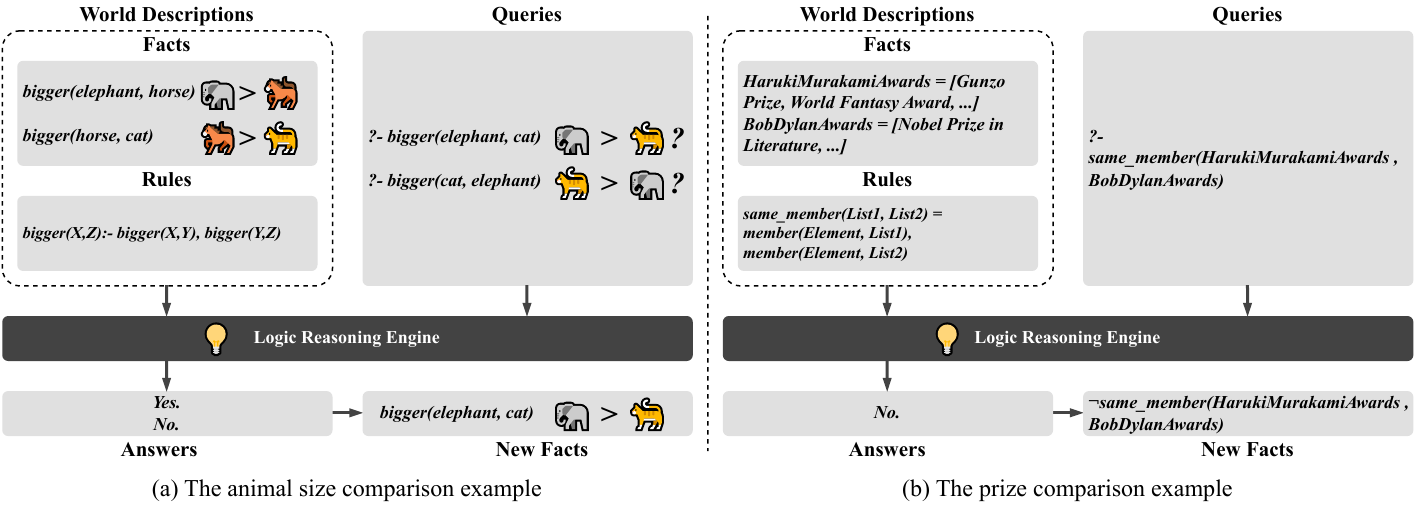}\\
    \caption{Examples of Logic Programming.}
    \label{fig:lg-example}
\end{figure}

Some existing works~\cite{Ye-Et-Al:2023:SAT, pan-etal-2023-logic, olausson-etal-2023-linc} have already integrated logical programming with large language models in an attempt to enhance their logical reasoning capabilities. In this work, we focus on leveraging logical programming to automate the testing of hallucinations in LLMs.
Logic programming languages are declarative, i.e., programming with these languages means describing the world. 
Using the programs means asking questions about the previously described world.
Based on the answers to the questions from the logic reasoning engine, according to the world description, we can derive new facts.
\autoref{fig:lg-example} shows an example of how logic programming works. 

Logic programming allows the programmer to specify the rules and facts, enabling the Prolog interpreter to infer answers to the given queries automatically. 
Here we explain some key concepts:

\head{Program.}
A Prolog program consists of two parts: a list of facts ($\widetilde{\relation}$) and a list of rules ($\widetilde{\drule}$). 
Throughout the paper, we use the over-tilde notation to denote a list of items. For example, $\widetilde{\entity}$ refers to a list of entities, i.e., $\entity_1, \dots, \entity_n$. 

\begin{equation}
\begin{aligned}
\m{(Program)} & \quad  \Prolog &{  ::=  } & \quad  \widetilde{\relation}\,{+}{+}\, \widetilde{\drule}
\end{aligned}
\label{eq:program}
\end{equation}

\head{Facts.}
A fact is a statement defining a relation as being true.
It is made up of a $predicate$ and several $entities$.
It is denoted as:
\begin{equation}
\begin{aligned}
\m{(Predicate)} & \quad  \relation &{  ::=  } & \quad  \m{\nm}\,(\widetilde{\entity})
\end{aligned}
\label{eq:fact}
\end{equation}
An example is $bigger(horse, cat)$, which means horses are bigger than cats.
Another example is $member(Gunzo Prize, HarukiMurakamiAwards)$, which means that the Gunzon Prize is in the list of prizes awarded to Haruki Murakami.

\head{Rules.} 
A Prolog rule is a Horn clause that comprises a head predicate and a list of body predicates placed on the left and right side of the arrow symbol ($\hornarrow$).
A rule means that the left-hand side is logically implied by the right-hand side. The rule bodies are either positive or negative relations, corresponding to the requirements upon the presence or absence of facts. 
We use ``$\relation$'' and 
``$\neg\,\relation$'' as abbreviations for  
``${\tt{Pos}}~\relation$'' and ``${\tt{Neg}}~\relation$'', respectively.
It is denoted as:
\begin{equation}
\begin{aligned}
\m{(Rule)}  & \quad  \m{\drule} &{ ::=  } & \quad 
\relation ~\hornarrow~ \widetilde{body}
\\ 
\m{(Rule~Bodies)}  & \quad  ~~\m{body} &{  ::=  } & \quad 
{\tt{Pos}}~ \relation
\,\mid\, {\tt{Neg}}~ \relation 
\end{aligned}
\label{eq:rule}
\end{equation}
An example is $\m{bigger(X,Z)\hornarrow bigger(X,Y), bigger(Y,Z)}$, which means the $bigger$ relation is \textbf{transitive}.
Another example is $\m{smaller(X,Y) \hornarrow bigger(Y,X)}$, which means $smaller$ is an \textbf{inverse} relation of $bigger$.
The last example here is $\m{same\_member(List1, List2)}$, which is $true$ if there exists at least one $Element$ that is a member of both $List1$ and $List2$.
It is a \textbf{composite} type of two $member$ predicates.

\head{Queries.} 
A query has the same structure as the body of a rule, i.e., it is a sequence of predicates separated by commas and executed against a database of facts. 
The logic reasoning engine will answer $Yes$ if the sequence of predicates in the query is $True$ according to the facts and rules.
Otherwise, it will answer $No$.

An example query is $\m{\text{?}\,\text{-}\,bigger(elephant, cat)}$, which means asking the logic reasoning engine whether elephants are bigger than cats.
Another example is $\text{?}\,\text{-}\,same\_member(HarukiMurakami\\Awards, BobDylanAwards)$, which means asking if the awards won by Haruki Murakami and Bob Dylan have overlapped.

\head{Reasoning Rules.}
As shown in \autoref{fig:lg-example}, generating new facts through logic programming requires facts (\autoref{eq:fact}), rules (\autoref{eq:rule}), queries, and answers to the queries.
To simplify the notation of this process, we bring up the concept of \textit{reasoning rules} in this paper, which describes the inference process of using facts and rules (predicates) to reach the conclusion (a new fact in the form of a predicate) by omitting the process of querying and analyzing the query answers.
A reasoning rule is denoted in this form:
\begin{equation}
\begin{aligned}
\frac{
        \begin{array}{c}
          \relation_1, \relation_2, ...\\
        \end{array}
    }{
    \begin{array}{c}
     \relation_{\m{new}}  
      \end{array} 
      }
\end{aligned}
\label{eq:reasoning rules}
\end{equation}

\section{Motivating Example}\label{sec:motivating}

\begin{figure}[!ht]
    \centering
    \includegraphics[width=\linewidth]{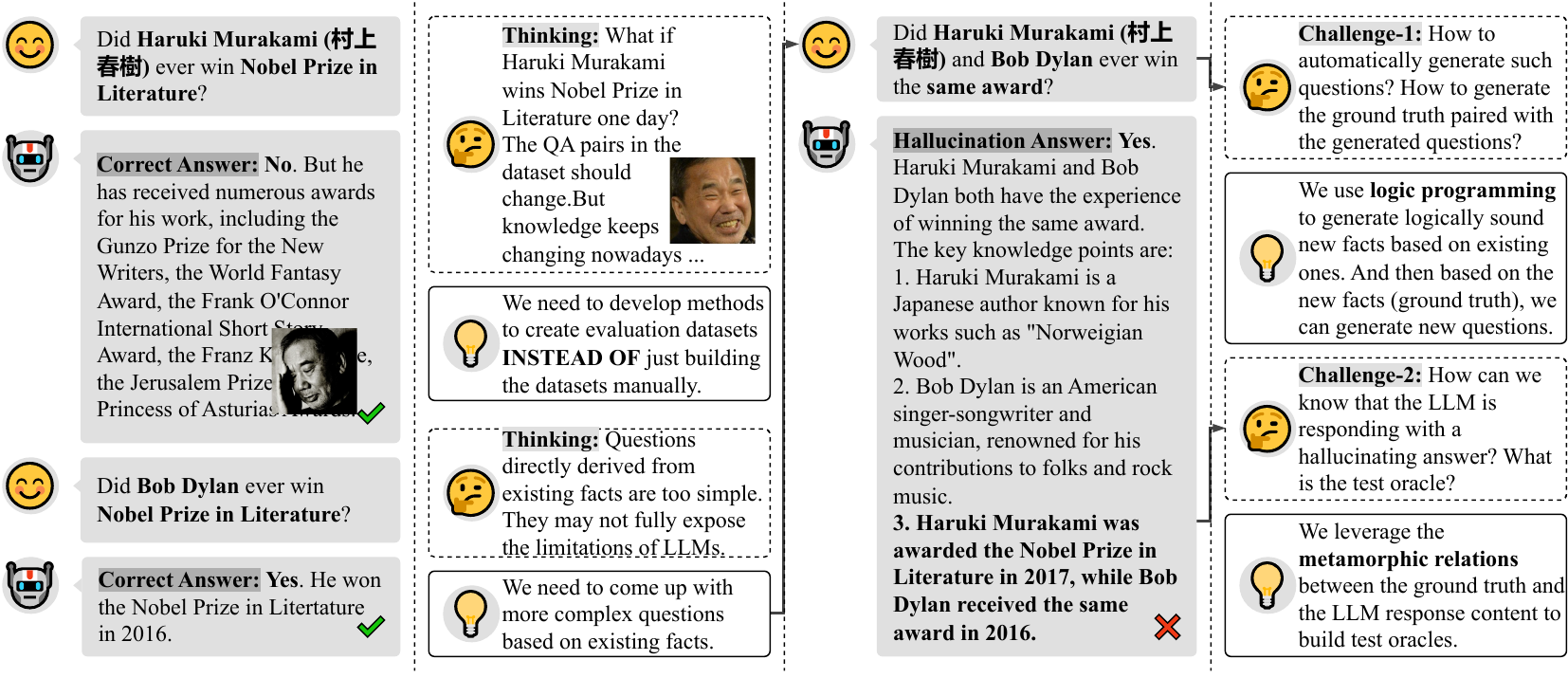}\\
    \caption{Motivating Example.}
    \label{fig:motivating}
\end{figure}

\autoref{fig:motivating} shows a motivating example of \tool{}.
Assume we have the facts about whether Haruki Murakami and Bob Dylan have won the Nobel Prize, as illustrated in the left sub-figure.
The question to ask LLMs is straightforward: We can ask whether Haruki Murakami/Bob Dylan has won the Nobel Prize or not.
Asking and verifying this knowledge requires no logic reasoning.
However, the straightforward questions are often not enough to unveil hallucinations.
\textbf{Therefore, more diversified questions (questions with intertwined and complex information, as illustrated in the right sub-figure) are needed.}

In order to generate more diversified benchmarks, previous research~\cite{yu2023kola, HaluEval} involves human experts to generate the questions and annotate the answers for hallucination checking.
Although the manually generated benchmarks can unveil certain hallucinations, they suffer from several drawbacks.
\textbf{The landscape of knowledge is dynamic, with new information continuously surfacing and older information becoming obsolete.} If facts change continuously over time, for instance Haruki Murakami were to win the Nobel Prize in the future, this would necessitate regular updates and corrections to the ground truth in existing datasets to reflect them. However, maintaining the accuracy of these benchmarks demands a significant amount of manual labor.
Additionally, the quality of the questions might be inconsistent due to the differences in the experience and skills of the human experts who create them.
Consequently, the efficiency and soundness of the manually generated benchmarks are not guaranteed.

The limitations of the manually generated benchmarks motivate the need for an automated technique to test for hallucinations in LLMs.
Nevertheless, automatically generating diverse benchmarks is challenging.
\textbf{First, generating suitable and valid questions is challenging~(challenge\#1).}
While it is important for the questions in the testing benchmark to cover a diverse range of scenarios, they cannot be randomly generated or arbitrarily selected. Instead, the questions must be logically coherent and aligned with well-established factual knowledge and ground truth.
\textbf{Second, deriving the test oracles for detecting hallucinations is challenging~(challenge\#2).} The LLM's answer is typically expressed in lengthy and potentially complex sentences. The key to determining if an LLM has produced an FCH lies in assessing whether the overall logical reasoning behind its answer is consistent with the established ground truth. Automatically analyzing and comparing the intricate logical structures within the LLM's response and the factual ground truth remains an inherently difficult task.

These two challenges can both be addressed by leveraging logic programming.
We can derive new logically sound facts based on existing knowledge.
With the new facts, we can then generate diverse questions and their ground truth answers.
With the ground truth answers, we can generate test oracles to capture hallucinations.
In short, the idea of using logic programming to tackle the challenges motivates the design of \tool{}.

\section{Methodology}\label{method}

\begin{figure*}[!ht]
    \centering
    \includegraphics[width=0.95\textwidth]{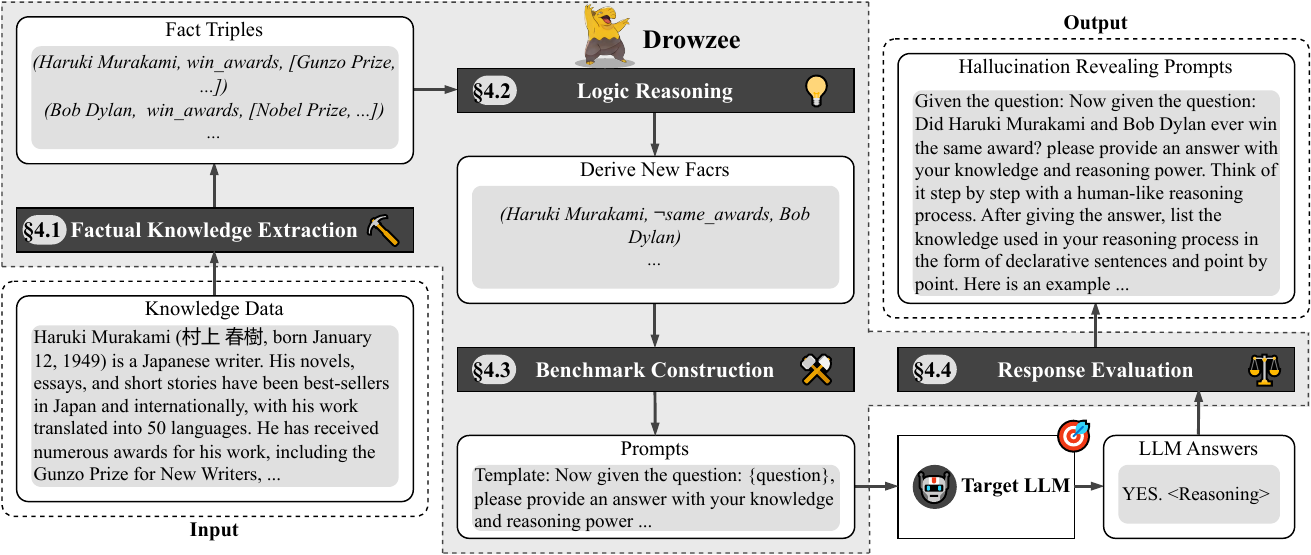}\\
     \caption{The Workflow of \tool.~*{\footnotesize The avatar is painted by one of the authors to avoid copyright issues.}}
    \vspace{-0.3cm}
    \label{fig:methodology}
\end{figure*}

We design and implement \tool{} to address the aforementioned challenges, the workflow of which is illustrated in \autoref{fig:methodology}. \tool{} is comprised of the following four modules, with each module to be detailed later. 
\begin{itemize}[leftmargin=*]
\item \textbf{Factual Knowledge Extraction (\S\ref{knowledge}):} Based on voluminous knowledge database dumps, \tool acquires fundamental information and factual triples of valid entities.

\item \textbf{Logical Reasoning (\S\ref{logic}):} In this module, \tool leverages reasoning rules to generate sound and diverse facts as new ground truth knowledge. 

\item \textbf{Benchmark Construction (\S\ref{prompt}):} This module focuses on creating high-quality test case-oracle pairs from the newly-derived ground truth knowledge. 
The test oracles are generated based on a simple yet effective metamorphic relation:
\emph{Since the newly generated knowledge is sound, the questions complying with the knowledge should be answered with ``YES'' and the questions contravening the knowledge should be answered with ``NO''.}
This module also includes strategies for effectively and reliably generating or selecting prompts for interaction with LLMs. 

\item \textbf{Response Evaluation (\S\ref{response}):} The final module evaluates the responses from the LLMs and detects factual consistency automatically. It first parses LLM outputs using NLP to construct semantic-aware structures, then evaluates their semantic similarity to ground truth. Subsequently, it develops similarity-based oracles applying metamorphic testing to assess consistency between LLM responses and ground truth.
\end{itemize}

\subsection{Factual Knowledge Extraction}\label{knowledge}
This step aims to extract fundamental facts from the input knowledge data into fact triples that can be utilized for logical reasoning. 

Existing knowledge databases~\cite{freebase, DBpedia, Yago, WordNet} not only encompass a vast array of documents and pages but also provide available structured data. Extracted from knowledge databases, the structured data would become an ideal resource for the construction and enrichment of factual knowledge. Thus, the genesis of our test case data is exclusively rooted in the entities and structured information sourced from current knowledge databases, ensuring a sophisticated and well-informed foundation for our testing framework. Basically, we follow the categorization of entities and relations used by WikiPedia~\cite{DBpedia} to perform the identification. Figure~\ref{table:categories} shows the categories of the entities.
Figure~\ref{table:relations} shows the categories of the relations and some example fact triples. 

The detailed process is outlined in Algorithm~\ref{alg:ground_truth}. As defined in the previous Equation~\ref{eq:fact}, we extract the facts in the structure of three-element predicates, i.e.,  $\nm\,(\Subj,\Obj)$, where ``$\Subj$'' (stands for $\m{subject}$) and ``$\Obj$'' (stands for $\m{object}$) are entities, and ``$\nm$'' is the name of the predicate.
The facts extraction is done on a per-category basis, implementing a divide-and-conquer strategy, which efficiently integrates all the facts ranging from all the categories. 
As shown in \algoref{alg:ground_truth}, for any given entity category and relation category, the function  
$\textsc{ExtractGroundFacts}$
iterates through all possible entities and relations. For each combination ($\m{entity}, \nm$), it queries the database using the $\textsc{QueryDB}$ function (Lines 3-6), which retrieves all three-element facts established with the specific predicate $\nm$ and the argument $\m{entity}$ placed either in the subject or the object position.
\begin{figure}[htbp]
    \centering
    \begin{subfigure}[b]{0.49\linewidth}
        \setlength{\tabcolsep}{1ex}
    	\resizebox{\linewidth}{!}{
    	\begin{tabular}{l l c}
        \toprule 
        \textbf{Category Type} & \textbf{Description}\\
        \midrule
        \textbf{Culture and the Arts} & Famous films, books, etc.\\ 
        \midrule
        \textbf{Geography and Places} & Countries, cities and locations. \\
        \midrule
        \textbf{Health and Fitness} & Diseases and disease-causing genes. \\
        \midrule
        \textbf{History and Events} & Famous historical events, etc. \\
        \midrule
        \textbf{People and Self} & Important figures and contributors. \\
        \midrule
        \textbf{Mathematics and Logic} & Common formulas and theorems. \\
        \midrule
        \textbf{Natural and Physical Sciences} & Celestial bodies and astronomy. \\
        \midrule
        \textbf{Society and Social Sciences} & Major social institutions, etc.\\ 
        \midrule
        \textbf{Technology and Applied Sciences} & Computer science, etc. \\
        \bottomrule 
    \end{tabular}}
    	\caption{Entity Categorization.}
            \label{table:categories}
    \end{subfigure}
    \hfill
    \begin{subfigure}[b]{0.49\linewidth}
        \setlength{\tabcolsep}{1ex}
    	\resizebox{\linewidth}{!}{
    	\begin{tabular}{l l}
        \toprule 
        \textbf{Category Type} & \textbf{Example}\\
        \midrule
        \textbf{Noun Phrase} & \begin{tabular}[l]{@{}l@{}} \textit{place\_of\_birth\,(Barack Obama, Honolulu).}\\ \textit{genre\,(28 Days Later, horror film).} \end{tabular}\\
        \midrule
        \textbf{Verb Phrase in Passive Voice} & \begin{tabular}[l]{@{}l@{}} \textit{killed\_by}\,\textit{(John F. Kennedy, Lee Harvey Oswald)}.\\ \textit{located\_in\_time\_zone\,(Arizona, UTC-07:00).}\\ 
        \end{tabular}\\
        \midrule
        \textbf{Verb Phrase in Active Voice} & \begin{tabular}[l]{@{}l@{}} \textit{follows\,(4769 Castalia, 4768 Hartley).}\\ \textit{replaces\,(American Broadcasting Company,} \\ \qquad \, \textit{NBC Blue Network).} \end{tabular}\\
        \bottomrule 
    \end{tabular}}
    	\caption{Relation Categorization.}
            \label{table:relations}
    \end{subfigure}
    \caption{Entity and Relation Categorization.}
\end{figure}

\begin{algorithm}[!ht]
\caption{Facts Extraction}
\label{alg:ground_truth}
\small
\begin{algorithmic}[1]
\Require 
Entity Category (\entityCat), Relation Category (\relationCat)
\Ensure Ground Facts ($\groundTruthTriples$)
\Function{ExtractGroundFacts}{
\entityCat, \relationCat}
    \State $\groundTruthTriples \gets []$ \Comment{\commentstyle{Initialization}}
    \For{~$\m{entity}$ $\in$ \entityCat~} \Comment{\commentstyle{Iterate over each entity}}
        \For{~$\nm$ $\in$  \relationCat~} \Comment{\commentstyle{Iterate over each relation}}
            \State $\widetilde{\relation} \gets$ \Call{QueryDB}{$\m{entity}$, $\nm$} 
            \Comment{\commentstyle{Retrive ground facts}}
            \State $\groundTruthTriples.\m{append}(\widetilde{\relation})$ \Comment{\commentstyle{Extend the ground facts}}
        \EndFor
    \EndFor
    \State \Return $\groundTruthTriples$ \Comment{\commentstyle{Return the ground facts}}
\EndFunction
\end{algorithmic}
\end{algorithm}

\subsection{Logical Reasoning}\label{logic}
This step aims to derive additional, enriched information from previously extracted factual knowledge. 
\tool uses a logical programming-based processor to automatically generate new factual knowledge. 
This allows us to take one or more factual knowledge triples as input and generate a derived triple as output with five types of inference rules.
 
To tackle the primary concern of generating FCH test cases with variability, we design five types of reasoning rules (\autoref{eq:reasoning rules}) prevalently adopted in several literature~\cite{zhou2019completing, ren2020beta, liang2022reasoning, TIAN2022100159, abboud2020boxe} in the context of knowledge reasoning. This provides sound strategies to prepare new facts for further test case generation.
\tool will exhaustively apply all the rules to all their relevant fact triples to generate new knowledge.
The definitions of the five types of rules are detailed as follows.

\noindent\textbf{\textit{Rule\#1: Negation Reasoning.}} Based on a given factual knowledge, we can determine whether the opposite of this fact is correct or incorrect by applying Definition~\ref{def:neg}.
\begin{definition}\label{def:neg}
\textbf{Negation Reasoning Rule $[Neg]$.} Given a factual knowledge triple $(s, \nm, o)$, then we can derive the new triple $(s, \overline{\nm}, o)$ is not valid. $\overline{\nm}$ indicates the negation of the relation $\nm$.
\end{definition}

\[
    \frac{
        \begin{array}{c}
          \nm(s, o)\\
        \end{array}
    }{
    \begin{array}{c}
     \neg ~ \overline{\nm}(s, o)   
      \end{array} 
      }
    [Neg]
\]
An example of this type of rule is: $\frac{
        \begin{array}{c}
          \m{was}(s, o)\\
        \end{array}
    }{
    \begin{array}{c}
     \m{\neg ~ wasn't}(s, o)   
      \end{array} 
      }
    [Neg]$.
    
With this rule, from the triple \emph{(Haruki Murakami, won, the Nobel Prize in Literature in 2016)}, we derive that the negation of this triple \emph{(Haruki Murakami, did not win, the Nobel Prize in Literature in 2016)} contains false factual knowledge. 

\noindent\textbf{\textit{Rule\#2: Symmetric Reasoning.}} In symmetric relations, if the subject and object in a triple maintain coherence upon interchange, a new triple can be deduced in accordance with Definition~\ref{def:symmetric}.
\begin{definition}\label{def:symmetric}
\textbf{Symmetric Reasoning Rule $[Sym]$.} Given a factual knowledge triple $(s, \nm, o)$, then we can derive a new triple $(o, \nm, s)$.
\end{definition}

\[
    \frac{
        \begin{array}{c}
          \nm(s, o)\\
        \end{array}
    }{
    \begin{array}{c}
      \nm(o, s)   
      \end{array} 
      }
    [Sym]
\]

An example of this type of rule is: $\frac{
        \begin{array}{c}
          \m{different\_from}(s, o)\\
        \end{array}
    }{
    \begin{array}{c}
      \m{different\_from}(o, s)   
      \end{array} 
      }
    [Sym]$.

With this rule, from the original triple \emph{(Haruki Murakami, different\_from, Haruki Uemura)}, we derive a new triple \emph{(Haruki Uemura, different\_from, Haruki Murakami)} (Haruki Uemura is a Japanese judoka). Note that the symmetric reasoning rule is primarily utilized within the composition reasoning rule~(to be detailed next) and does not introduce new knowledge on its own.

\noindent\textbf{\textit{Rule\#3: Inverse Reasoning.}} In an inverse relation, the subject and object can be reversely linked through a variant of the original relation, as defined in Definition~\ref{def:inverse}.
\begin{definition}\label{def:inverse}
\textbf{Inverse Reasoning Rule $[Inverse]$.} Given a factual knowledge triple $(s, \nm, o)$ and a reversed relation $\nm'$ of $R$, then we can derive a new triple $(o, \nm', s)$.
\end{definition}

\[
    \frac{
        \begin{array}{c}
          \nm(s, o), \nm'=Reverse(\nm)\\
        \end{array}
    }{
    \begin{array}{c}
      \nm'(o, s)   
      \end{array} 
      }
    [Inverse]
\]

An example of this type of rule is: $\frac{
        \begin{array}{c}
          \m{influence\_by}(s, o)
        \end{array}
    }{
    \begin{array}{c}
      \m{influence}(o, s)   
      \end{array} 
      }
    [Inverse]
    $.
With this rule, from the triple \emph{(Haruki Murakami, influence\_by, Richard Brautigan)}, we can derive a new triple \emph{(Richard Brautigan, influence, Haruki Murakami)}.

\noindent\textbf{\textit{Rule\#4: Transitive Reasoning.}} In transitive relations, if the object in one triple is the subject of the second triple, we can therefore derive a new triple following the Definition~\ref{def:transitive}.
\begin{definition}\label{def:transitive}
\textbf{Transitive Reasoning Rules $[Trans]$.} Given two factual knowledge triples $(s_1, \nm, o_1)$ and $(s_2, \nm, o_2)$, if $o_1$ is semantically equivalent to $s_2$, then we can derive a new triple $(s_1, \nm, o_2)$.
\end{definition}

\[
    \frac{
        \begin{array}{c}
          \nm(s_1, o_1),~ \nm(s_2, o_2), ~o_1 = s_2\\
        \end{array}
    }{
    \begin{array}{c}
      \nm(s_1, o_2)   
      \end{array} 
      }
    [Trans]
\]
An example here is: 
\[
    \frac{
        \begin{array}{c}
          loc\_in(s_1, o_1),~ loc\_in(s_2, o_2), ~o_1 = s_2\\
        \end{array}
    }{
    \begin{array}{c}
      loc\_in(s_1, o_2)   
      \end{array} 
      }
    [Trans].
\]
With this rule, from triples \emph{(Haruki Murakami, locate\_in, Kyoto)} and \emph{(Kyoto, locate\_in, Japan)}, we derive a new triple \emph{(Haruki Murakami, locate\_in, Japan)}.

\noindent\textbf{\textit{Rule\#5: Composite Reasoning.}} The previous four reasoning rules are all meta-rules capturing the most basic and fundamental logical relations among the facts and rules. 
Several basic reasoning rules can be chained together to form a composition reasoning rule if the relations in the rules have logical relations.
Composite reasoning rules can generate knowledge that requires multiple steps of reasoning.

\begin{definition}\label{def:composite}
\textbf{Composite Reasoning Rules $[Comp]$.} Given multiple basic reasoning rules or predicates $[Rule_i] \in \{[Neg],[Sym],[Inverse],[Trans],[Predicates]\}$, we can chain them up to form a new composite reasoning rule.
\end{definition}

\[ 
   \frac{
    \frac{
        \frac{
        \begin{array}{c}
         \nm_{1\_Rule_1}(...), \nm_{2\_Rule_1}(...), ...\\
          \end{array} 
        }{
        \begin{array}{c}
          R_1
          \end{array} 
          }
        [Rule_1],\; ...
    }
    {
    \frac{\frac{
        \begin{array}{c}
          ...
          \end{array} 
        }{
        \begin{array}{c}
          ...
          \end{array} 
          }
        [...],\;...
    }{
        \frac{
        \begin{array}{c}
          \nm_{1\_Rule_i}(...), \nm_{2\_Rule_i}(...), ...\\
          \end{array} 
        }{
        \begin{array}{c}
         R_{\m{i}} 
          \end{array} 
          }
        [Rule_i], \;  ...
    }
   }}{
        \begin{array}{c}
          R_{\m{new}} 
          \end{array} 
   }
    [Comp]
\]

   

The process of applying these various rules to the ground truth triples extracted in the previous module can be referenced in Algorithm~\ref{alg:logic_reasoning}. An automatic rule generator could be designed at the first stage to iterate its predicates and generate the derivation rules $\drule$ according to the relation category (as in Lines 3-4). 
The corresponding query problems are also generated and mapped to the generated rules, which could be applied to the Prolog query later. 
With the predetermined rules, we can be assisted with the Prolog engine, asserting all the related triples and consulting the reasoning rules, as outlined in Lines 5-6. We use $\llbracket \relation \rrbracket_{\Prolog}$ to denote the query results of $\relation$ w.r.t the Prolog program $\Prolog$. 
When $\relation$ contains no variables,
it returns Boolean results indicating the presence of the fact; otherwise, it outputs all the possible instantiations of the variables. 
Then as stated in Lines 7-9, by obtaining solutions from Prolog, we can generate new knowledge triples based on the entities and their relations provided. For each instantiation that contains one subject ``s'' and one object ``o'', we then compose them with the new predicate, which is taken as one derived fact.  
These derived facts are later used to generate test cases.

\begin{algorithm}[!ht]
\caption{Deriving New Facts}
\label{alg:logic_reasoning}
\small
\begin{algorithmic}[1]
\Require Ground Facts ($\groundTruthTriples$), Relation Category (\relationCat)
\Ensure Derived Facts ($\derivedFacts$)
\Function{DerivingFacts}{$\groundTruthTriples$, \relationCat}
\State $\derivedFacts \gets []$ \Comment{\commentstyle{Initialization}}
\For{$\nm$ in \relationCat}
\Comment{\commentstyle{Iterate each predicate}}
\State $\deriveRules{\nm}{\nm_{\m{new}}}{{\drule}}$\Comment{\commentstyle{Obtain the reasoning rule, and the new predicate}}
\State $\Prolog \gets \groundTruthTriples{+}{+}{\drule}$ \Comment{\commentstyle{Construct the Prolog program}} 
\State $\m{instantiations} \gets \llbracket \nm_{\m{new}}(\SUBJ, \OBJ)\rrbracket_{\Prolog}$ 
\Comment{\commentstyle{Obtain concrete entities}}
\For{(\Subj, \Obj) in $\m{instantiations}$}
\Comment{\commentstyle{Iterate each entity tuple}}
\State $\relation_{\m{new}} \gets \nm_{\m{new}}(\Subj, \Obj)$ 
\Comment{\commentstyle{Construct the derived fact}}
\State $\derivedFacts.\m{append}(\relation_{\m{new}})$ \Comment{\commentstyle{Append the derived facts}}
\EndFor
\EndFor
\State \Return $\derivedFacts$ \Comment{\commentstyle{Return the derived facts}}
\EndFunction 
\end{algorithmic}
\end{algorithm}

\vspace{-0.5cm}
\subsection{Benchmark Construction}\label{prompt}
From the derived triples, this module outlines our approach to constructing question-answer~(Q\&A) pairs and prompts to facilitate the automatic testing of FCH. 

In addressing the obstacle of high human effort demanded in the test oracle generation process, we design an automated generation of test case-oracle pairs based on mapping relations between various entities to problem templates, greatly reducing reliance on manual effort.

\head{Question Generation.}
To ensure effective and systematic test cases and prompt generation, we have adopted a method that utilizes entity relations mapping to predefined Q\&A templates. 
In the construction of relation-based Q\&A templates, one key aspect lies in aligning various types of relations with the corresponding question templates from the derived triples, i.e., the predicate type in the triple. Different relation types possess unique characteristics and expressive requirements, leading to various predefined templates. 
As listed in Table~\ref{table:template}, we map the relation types to question templates based on speech and the grammatical tense of the predicate, to guarantee comprehensive coverage. Beyond these universal templates, for hard-to-describe predicates, we employ customized templates to generate valid Q\&A pairs. 
To enhance the construction of natural language formatted questions, we also leverage the LLM to refine the structure of  Q\&A pairs.

Another key aspect is regarding the automatic and reliable answer generation. We note that the answer to the corresponding question is readily attainable from the factual knowledge in the triple. 
Primarily, it is easy to determine whether the answer is true/false based on the derived triples. Meanwhile, mutated templates with positive and negative semantics via the usage of synonyms or antonyms, which greatly enhance the question diversity, can be treated in a similar manner as the negation rule defined in Section~\ref{logic}. 
Specifically, if the answer to a question with original semantics is Yes/No, then for a question with mutated opposite semantics, the corresponding answer would naturally be the opposite, i.e., No/Yes. For example, after obtaining the original Q\&A pair \textit{- Is it true that Crohn's disease and Huntington's disease could share similar symptoms and signs? - Yes.}, we can use some antonyms to mutate it into \textit{- Is it true that Crohn's disease and Huntington's disease have totally different symptoms and signs? - No.}

\begin{table}[!ht]
\setlength{\tabcolsep}{1pt}
\centering
\caption{Relation-Template Mapping Pattern.}
 \vspace{-2mm}
\label{table:template}
\resizebox{0.8\linewidth}{!}{
\begin{tabular}{l l}
\toprule 
\textbf{Relation} & \textbf{Template Examples}  \\
    \midrule
{Noun Phrase} & \begin{tabular}[l]{@{}l@{}} - Is it true that 
$\langle \m{Subject}\rangle$ and 
$\langle\m{Object}\rangle$ share 
$\langle\m{Relation}\rangle$? 
\\ - $\langle\m{Subject}\rangle$ and $\langle\m{Object}\rangle$ have/made/shared totally different $\langle\m{Relation}\rangle$. \\
\ \ \, Please judge the truthfulness of this statement.
    \end{tabular}  \\
    \midrule
    \begin{tabular}[l]{@{}l@{}} Verb Phrase \\ (Passive Voice) \end{tabular} & \begin{tabular}[l]{@{}l@{}} - Is it true that $\langle \m{Subject}\rangle$ is/was/are/were $\langle \m{Subject}\rangle$ $\langle\m{Object}\rangle$? \\ - It is impossible for $\langle \m{Subject}\rangle$ to be $\langle\m{Relation}\rangle$ $\langle\m{Object}\rangle$. Am I right?
    \end{tabular}  \\
    \midrule
\begin{tabular}[l]{@{}l@{}} Verb Phrase \\ (Active Voice) \end{tabular}
 & \begin{tabular}[l]{@{}l@{}} - Is it true that 
 $\langle \m{Subject}\rangle$
 $\langle\m{Relation}\rangle$
 $\langle\m{Object}\rangle$?  \\ - $\langle \m{Subject}\rangle$ $\langle\m{Relation}\rangle$ $\langle\m{Object}\rangle$. 
 \end{tabular}  \\
    \bottomrule 
\end{tabular}
}
\end{table}
\begin{table}[!ht]
    \setlength{\tabcolsep}{1ex}
	\centering
    \large 
	\caption{Prompt Template.}
        \label{table:prompt}
        \vspace{-0.3cm}
	\resizebox{\linewidth}{!}{
	\begin{tabular}{l}
    \toprule 
    \rowcolor{mycolor}
    \textbf{\instruction:} Answer the question with your knowledge and reasoning power.\\
    \midrule
    \rowcolor{mycolor} \textbf{\query:} Now given the question: \textit{question}, please provide an answer with your knowledge and reasoning power.\\ 
    \rowcolor{mycolor} Think of it step by step with a human-like reasoning process.\\
    \rowcolor{mycolor} After giving the answer, list the knowledge used in your reasoning process in the form of declarative sentences and point by point.\\
    \rowcolor{mycolor} Here is an example. Question: During Barack Obama held the position as the president of the USA, were any films directed by \\
    \rowcolor{mycolor}James Cameron released? \\
    \rowcolor{mycolor} Supposed Response: Yes, during Barack Obama's presidency from 2009 to 2017, one film directed by James Cameron was released\\
    \rowcolor{mycolor}- Avatar in 2009. \\
    \rowcolor{mycolor} The key knowledge points used in this reasoning process are:\\
    \rowcolor{mycolor} 1. Barack Obama was the US President from January 20, 2009 to January 20, 2017.\\
    \rowcolor{mycolor} 2. James Cameron is a famous film director known for movies like Titanic, Avatar, Terminator 2, etc. \\
    \rowcolor{mycolor} 3. Cameron's only film release during Obama's presidency was Avatar in 2009.\\
    \bottomrule 
\end{tabular}}
\end{table}
\vspace{-0.3cm}
\head{Prompt Construction.}
As illustrated in Table~\ref{table:prompt}, before initiating our interaction with LLMs, we predefine specific instructions and prompts, requesting the model to utilize its inherent knowledge and inferential capabilities to deliver explicit (yes/no/I don't know) judgments on our queries. Additionally, we instruct the model to present its reasoning process in a template following the judgment. The primary aim is to ensure LLMs provide easily assessable responses by using standardized prompts and instructions. This approach also ensures that the model can exercise its reasoning abilities as effectively as possible under the given instructions and cues.

\subsection{Response Evaluation}\label{response}

The objective of our proposed module is to enhance the detection of FCH within LLM outputs, specifically focusing on the discrepancies between LLM responses and verified ground truth in Q\&A pairs. Recognizing the inherent challenges in directly accepting ``yes'' or ``no'' answers from LLMs due to potential inaccuracies, our approach underscores the importance of thoroughly analyzing the reasoning process presented by LLMs. This analysis is vital for accurately determining the factual consistency of LLM responses, thereby addressing the primary challenge in identifying FCH within LLM outputs.

To achieve automated detection of factual consistency, our methodology first incorporates a parsing step that leverages advanced NLP techniques. This step is designed to extract essential semantic elements from each sentence within LLM outputs, assembling these elements into a coherent, semantic-aware structure. The foundational premise of our approach is predicated on evaluating the semantic similarity between these constructed structures, aiming to discern the degree of consistency in their underlying semantics.
Subsequently, we propose the development of a list of similarity-based testing oracles. These oracles are instrumental in applying metamorphic testing principles, enabling us to systematically assess the consistency or inconsistency between LLM responses and the established ground truth. Note that our focus is on the accuracy of ground truth facts rather than highly specialized or sequential content like mathematical proofs. Consequently, during evaluation, we emphasize whether the entities and relations in the response align with the ground truth, regardless of the order in which the facts are presented. Our approach is structured around several critical steps, detailed as follows:

\textbf{Step 1. Preliminary Screening.} First, we eliminate scenarios in which the LLM declines to provide an answer, as indicated by the ``answer'' field of the LLM's response (as described in Algorithm~\ref{alg:re} Lines 3-4). Most of these responses arise because the LLM lacks the relevant knowledge for the reasoning process. Since these responses adhere to the LLM's principle of honesty, we classify them as correct and normal responses, denoted as $\m{CO}$ in the algorithm.

\textbf{Step 2. Response Parsing and Semantic Structure Construction.} As stated in Algorithm~\ref{alg:re} Lines 6-7, for the remaining suspicious responses, the \textsc{ExtractTriple} function is used to generate triples based on the statements contained in the \textit{reasoning process} part of the LLM's response. Then from the extracted triples ($\widetilde{\m{Trpl}}$), the \textsc{BuildGraph} function can construct a semantic structure $\m{SS_{resp}}$, where the $\m{entities}$ (i.e., the subject and object) are represented as nodes, and the $\m{relation}$ between them is illustrated as an edge connecting these nodes. Concurrently, the ground truth triples ($\rall$) associated with the question are used as input to construct a similar semantic structure $SS_{GT}$.

\textbf{Step 3. Similarity-based Metamorphic Testing and Oracles.} 
We apply metamorphic relations to detect and evaluate potential errors in LLM responses, based on the relationships between inputs and outputs, rather than relying on traditional labeled data. In our context, metamorphic relations specifically refer to comparing the similarity between semantic structures generated by LLMs and the ground truth counterparts, to identify and classify hallucination answers from LLMs (as mentioned in Algorithm~\ref{alg:re} Lines 8-17). Note that we provide four classifications: correct responses (denoted as $\m{CO}$), hallucinations caused by error inference ($\m{EI}$), hallucinations caused by erroneous knowledge ($\m{EK}$), and hallucinations containing both issues ($\m{OL}$).

\begin{algorithm}[!ht]
\caption{Response Evaluation}
\label{alg:re}
\small
\begin{algorithmic}[1]
\Require LLM Response ($\llmResponse$), All Ground Facts ($\rall$), Threshold ($\theta_{\m{e}}, \theta_{\m{n}}$)
\Ensure Evaluation Result Category~($CO, EK, EI, OL$)
\Function{EvaluateResponse}{$\llmResponse$, $\rall$, $\theta_{\m{e}}$, $\theta_{\m{n}}$}
    \State $CO, EK, EI, OL \gets []$ \Comment{\commentstyle{Initialization}}
    \If{$\llmResponse.answer = refusal$}
        \State
        $CO.\m{append}(\llmResponse)$ \Comment{\commentstyle{Preliminary Screening}}
    \Else
        \State $\widetilde{\m{Trpl}} \gets$ \Call{ExtractTriple}{$\m{Resp.reasoning}$} \Comment{\commentstyle{Extract useful triples}}
        \State $\semantic_{\m{resp}}, \semantic_{\m{ground}} \gets$ \Call{BuildGraph}{$\widetilde{\m{Trpl}}, \rall$} \Comment{\commentstyle{Build semantic structure}}
        \State $\m{s}_{\m{e}} \gets$ $\similarity_\m{e}${$(\semantic_{\m{resp}}$, $\semantic_{\m{ground}})$} \Comment{\commentstyle{Calculate edge similarity}}
        \State $\m{s}_{\m{n}} \gets$ $\similarity_\m{n}${$(\semantic_{\m{resp}}$, $\semantic_{\m{ground}})$} \Comment{\commentstyle{Calculate node similarity}}
        \If {$s_e < \theta_{e}$ and $s_n < \theta_{n}$}
            \State
            $OL.\m{append}(\llmResponse)$  \Comment{\commentstyle{Append hallucination related to both types}}
        \ElsIf{$\m{s}_{\m{e}} < \theta_{\m{e}}$}
            \State 
            $EI.\m{append}(\llmResponse)$  \Comment{\commentstyle{Append error inference hallucination}}
        \ElsIf{$\m{s}_{\m{n}} < \theta_{\m{n}}$}
            \State 
            $EK.\m{append}(\llmResponse)$  \Comment{\commentstyle{Append error knowledge hallucination}}
        \Else
            \State
            $CO.\m{append}(\llmResponse)$
            \Comment{\commentstyle{Append correct response}}
        \EndIf
    \EndIf
    \State \Return $CO, EK, EI, OL$ \Comment{\commentstyle{Return the categorized evaluation result}}
\EndFunction
\end{algorithmic}
\end{algorithm}

Specifically, the oracles for metamorphic testing can be divided into the following types:

\textbf{Edge Vector Metamorphic Oracle ($\m{MO_E}$)}: This oracle is based on the similarity of edge vectors between $\m{SS_{resp}}$ and $\m{SS_{ground}}$. If the vector similarity between the edges in the $\m{SS_{resp}}$ and those in $\m{SS_{ground}}$ falls below a predetermined threshold, it indicates that the LLM's answer significantly diverges from the ground truth, suggesting the presence of an FCH. Conversely, if the similarity meets or exceeds the threshold, the LLM's answer is considered to align with the ground truth. 
More specifically, we utilize Jaccard Similarity~\cite{J_S} to gauge the similarity score between edge vectors extracted from $\m{SS_{resp}}$ and those in $SS_{ground}$. 
$$\m{{J\_Sim}_E(SS_{resp}, SS_{ground}) = \frac{|\widetilde{E}_{\m{resp}} \cap \widetilde{E}_{\m{ground}}|}{|\m{\widetilde{E}_{resp}} \cup \m{\widetilde{E}_{ground}}|}},$$check if$$  \m{{J\_Sim}_E(SS_{resp}, SS_{ground})  < \theta_n} \ $$
where $\m{\widetilde{E}_{resp}}$ and $\m{\widetilde{E}_{ground}}$ denote the list of edges extracted from $\m{SS_{resp}}$ and $\m{SS_{ground}}$, and \( \theta_E \) is a predefined threshold~(to be detailed in Section~\ref{sec:ex_setup}). Intuitively, the similarity score is calculated as the proportion of identical edges shared between the two lists against the total number of unique edges in both lists. If the score is smaller than the threshold, then an FCH is detected. Note that when determining the joint and union of lists $\m{\widetilde{E}_{resp}}$ and $\m{\widetilde{E}_{ground}}$, we consider two edges as identical if their corresponding relations are identical or represented by synonymous words, and vice versa.  



\textbf{Node Vector Metamorphic Oracle ($\m{MO_N}$)}: This relation examines the similarity of node vectors between $\m{SS_{resp}}$ and $\m{SS_{ground}}$. 
Defined in a similar manner as $\m{MO_N}$, if the node similarity between the edges in the $\m{SS_{resp}}$ and those in $\m{SS_{ground}}$ falls below a predetermined threshold, it indicates that the LLM's answer significantly diverges from the ground truth, suggesting the presence of an FCH; vice versa.
$\m{MO_N}$ can be captured by the Jaccard Similarity, defined as follows:

$$\m{{J\_Sim}_N(SS_{resp}, SS_{ground}) = \frac{|\m{\widetilde{N}_{resp}} \cap \m{\widetilde{N}_{ground}}|}{|\m{\widetilde{N}_{resp}} \cup \m{\widetilde{N}_{ground}}|}},$$check if$$  \m{{J\_Sim}_N(SS_{resp}, SS_{ground})  < \theta_n} \ $$
where $\m{\widetilde{N}_{resp}}$ and $\m{\widetilde{N}_{ground}}$ denotes the list of nodes extracted from $\m{SS_{resp}}$ and $SS_{ground}$, and \( \theta_n \) is a predefined threshold~(to be detailed in Section~\ref{sec:ex_setup}). Intuitively, the similarity score is calculated as the proportion of identical nodes shared between the two lists against the total number of unique nodes in both lists. If the score is smaller than the threshold, then an FCH is detected. Note that when determining the joint and union of lists $\m{\widetilde{N}_{resp}}$ and $\m{\widetilde{N}_{ground}}$, we consider two nodes as identical if their corresponding entities are identical or represented by synonymous words, and vice versa.  

\section{Evaluation}\label{sec:eval}
Our evaluation targets the following research questions:

$\bullet$ \textbf{RQ1 (Effectiveness): How effective is \tool for identifying LLM FCH issues?} This RQ studies the effectiveness of \tool in generating test cases and identifying LLM FCH issues.

$\bullet$ \textbf{RQ2 (Hallucination Categorization and Analysis): What is the categorization of LLM FCH issues?}
This RQ categorizes the FCH issues of various LLMs identified by \tool. We also provide case studies for some specific cases, including temporal-related FCHs and out-of-distribution-data knowledge-related FCHs.

$\bullet$ \textbf{RQ3 (Comparison with Existing Works): How does \tool compare with existing approaches in detecting LLM FCH issues?} This RQ investigates whether \tool outperforms existing testing benchmarks and methods in constructing test cases and identifying LLM FCH issues. We conduct a qualitative analysis as well as a small-scale quantitative analysis of the accuracy of current hallucination detection methods compared with \tool.

$\bullet$ \textbf{RQ4 (Ablation Study): Whether the four types of logic reasoning rules can identify LLM FCH issues independently?} 
This RQ explores whether the logic reasoning rules of \tool can effectively identify LLM FCH issues separately.



\subsection{Experimental Setup}\label{sec:ex_setup}
\noindent \textbf{Knowledge Extraction.} We use Wikipedia and Wikidata as sources to extract entities and structured information as base factual knowledge. After downloading the latest Wikipedia dump, we employ wikiextractor~\cite{Wikiextractor2015} to extract relevant text from Wiki pages. In parallel, we invoke Wikidata's SPARQL~\cite{sparql} query module for the extraction of triples. 
Through data processing involving simplification and filtration, we amass a collection of basic factual knowledge, encompassing a sizeable number of 54,483 entities and 1,647,206 triples.

\noindent \textbf{Logic Reasoning Processor.} For the logic reasoning module, we apply SWI-Prolog~\cite{wielemaker2012swi}, an open-source advanced logical programming interpreter. To effectively prevent errors due to excessive stacked strings, and ensure the proper operation of the logical processor when inserting a large number of facts into Prolog, we employ a sampling method and extract a subset of entities to form a query. 

\noindent \textbf{Models Under Test.} To guarantee a reliable evaluation for RQ1 and RQ2, we evaluate six state-of-the-art large language models with \tool. Considering the diverse nature of LLMs, we select two distinct categories for in-depth analysis: the first category comprises API-accessible models with closed-source architecture including ChatGPT (gpt-3.5-turbo-0613) and GPT-4~\cite{OpenAI2023GPT4TR}, and the second category consists of open-source LLMs with deployability, including Llama2-7B-chat-hf, Llama2-70B-chat-hf~\cite{touvron2023llama}, Mistral-7B-Instruct-v0.2~\cite{jiang2023mistral}, and Mixtral-8x7B-Instruct-v0.1~\cite{jiang2024mixtral}. 

\noindent \textbf{Model Configuration.} We set the \textit{temperature} parameter to 0 to achieve more stable and conservative model outputs, ensuring consistency in the content generated during LLM testing. Additionally, we set the \textit{top-p} value to 0.9 and disable \textit{top-k} (set to 0) to filter out low-probability tokens and select the most likely tokens, thereby improving the accuracy of the generated results. To further validate the consistency of the LLM responses, we conducted significance tests to validate the consistency of the LLM responses. Specifically, we randomly selected 40 test cases from each domain under each rule, resulting in a total of 1440 test cases. Using GPT-3.5-turbo as an example, each test case is run five times under the previously described configuration to interact with the LLM. We then use the Sentence-bert model~\cite{reimers2019sentence} to calculate the pairwise cosine similarity between the five LLM responses generated for each test case. The consistency of the LLM responses is evaluated using Friedman tests~\cite{F_T}, a non-parametric statistical test commonly employed to detect differences in treatments across multiple test attempts. The results show no significant differences between the responses of different runs, with an approximate average p-value of 0.54. This confirms that the generated responses are consistent across multiple executions, justifying the use of a single run for evaluation purposes.

\noindent \textbf{Response Validation Threshold $\theta$.} To validate responses from LLMs as described in Section~\ref{response}, we apply StanfordOpenIE~\cite{angeli-etal-2015-leveraging, StanfordOpenIEWrapper} for knowledge triple extraction from LLM responses and then use Phrase-BERT~\cite{phrasebertwang2021} to calculate the vector similarity of nodes and edges from the constructed semantic structures. We also utilize GPT-4 to extract triples for some complex responses that StanfordOpenIE cannot handle effectively. Here we set the threshold to 0.8, considering knowledge triples as semantically equivalent if they exceed this threshold, and vice versa. To determine the threshold value, we sample 30 test cases and corresponding LLM responses from each of the nine knowledge domains listed in Figure~\ref{table:categories}. Through this analysis, we find that by setting the threshold values for both $\theta_E$ and $\theta_N$ at 0.8, with the given 270 test cases that are correctly classified, we can estimate the true positives among all test cases through \textit{Laplace's approach in the Sunrise problem}~\cite{laplace1951philosophical}, resulting in 99.6\% when detecting non-equivalent LLM answers as FCHs. In other words, all instances where an LLM's answer has a semantic similarity score below 0.8 compared to the ground truth were correctly identified as FCH cases. 

\noindent \textbf{Consistency of Words.} To ensure word consistency in experiments, we maintain several dictionaries. For symmetric relations, we have a dictionary that includes relations which retain their meaning when the subject and object are reversed. Additionally, we use synonym dictionaries provided by NLP libraries (e.g., WordNet~\cite{miller-1994-wordnet}) along with our own set of synonyms tailored for specific cases when validating the LLM responses.

\noindent \textbf{Running Environment.} Our experiments are conducted on a server running Ubuntu 22.04 with two 64-core AMD EPYC 7713, 512 GB RAM, and two NVIDIA A100 PCIe 80GB GPUs. Our experiments consume a total of 120 GPU hours.


\subsection{RQ1: Effectiveness}

To reveal the effectiveness of \tool, we evaluate the statistics of test cases generated by \tool and then evaluate the capabilities of identifying LLM FCH issues with the generated cases. 
To further assess the effectiveness of test cases for uncovering FCH issues in specific knowledge domains, we evaluate the performances of LLMs on test cases across various knowledge domains.

\begin{figure}[t]
    \centering
    \vspace{-0.3cm}
    \begin{subfigure}[b]{0.45\linewidth}
        \centering
        \includegraphics[width=\linewidth]{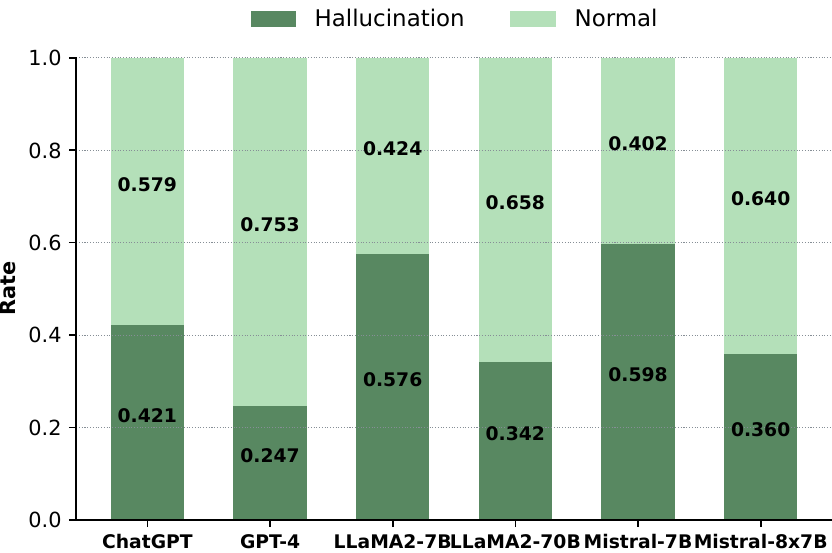}
        \caption{Overall Hallucination Rate of Various LLMs.}
        \label{fig:overall}
    \end{subfigure}
    \begin{subfigure}[b]{0.5\linewidth}
        \centering
        \includegraphics[width=\linewidth]{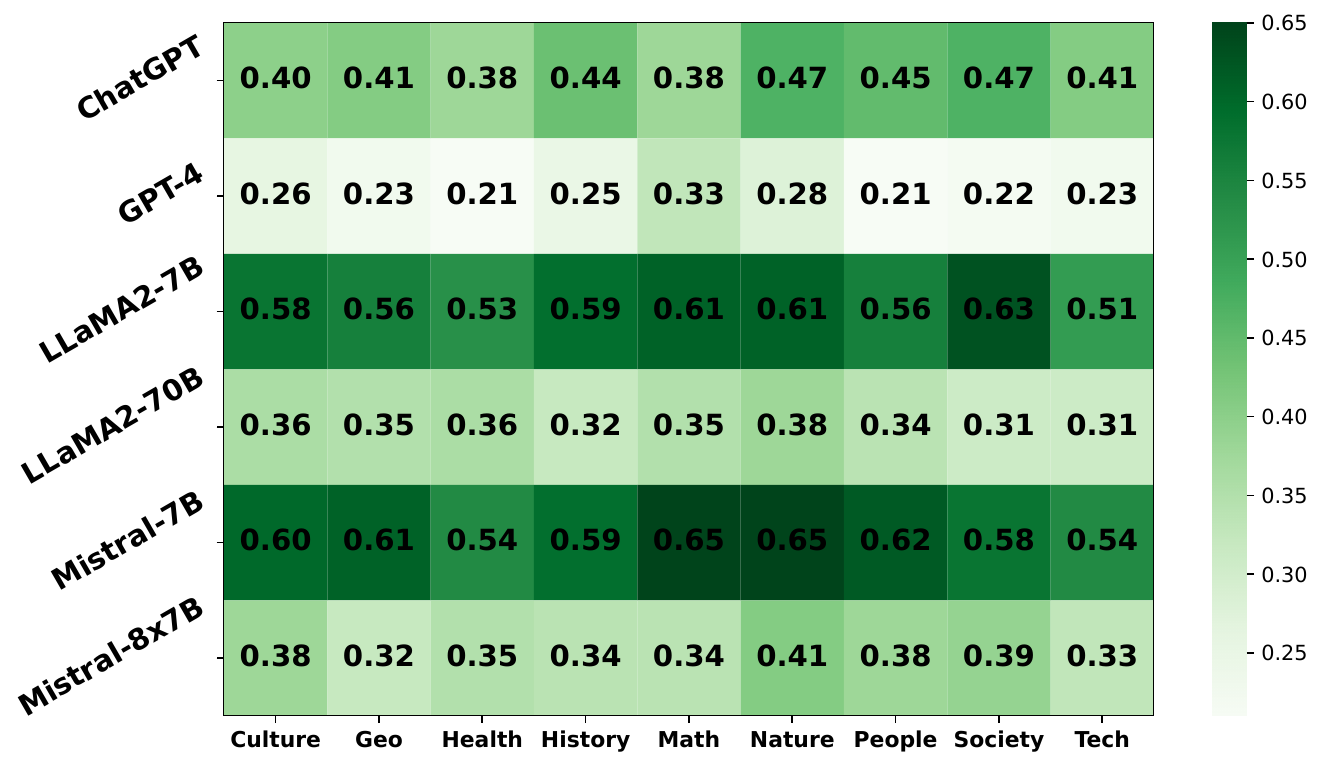}
        \caption{Hallucination Rate Heatmap of Specific Domain.}
        \label{fig:rq1.2}
    \end{subfigure}
    \caption{Effectivess of \tool.}
\end{figure}

\head{Effectiveness on Generating Q\&A Test Cases.} We apply \tool to generate a Q\&A test benchmark, amounting to a comprehensive total of 7,200 test cases, designed to provide a broad and detailed evaluation of LLM FCH issues across specific knowledge domains. 

\head{Effectiveness across LLMs.} We instruct LLMs under test utilizing Q\&A pairs derived from \tool and automatically label both hallucination and normal responses. Different LLMs might trigger different hallucinations on the same questions. The results are presented in Figure~\ref{fig:overall}, illustrating the proportion of FCHs versus normal responses from LLMs under test.

Among all models, GPT-4 exhibits the best performance, demonstrating the lowest proportion of hallucinatory responses in the test cases generated by \tool, at only 24.7\%, while ChatGPT falls slightly behind with 42.1\%. Open-source LLMs including Llama2-7B-chat-hf and Mistral-7B-Instruct-v0.2 with fewer parameters perform worse, but their counterparts with larger parameters (i.e., Llama2-70B-chat-hf and Mixtral-8x7B-Instruct-v0.1) achieve higher normal response rates surpassing ChatGPT on \tool. 
This indicates that the test cases generated by \tool successfully trigger hallucination responses across various LLMs when confronted with questions requiring logical reasoning capabilities.

\head{Effectiveness on Specific Domain Knowledge for Each LLM.}
To further explore the effectiveness of \tool in identifying FCH issues spanning various domains of LLMs, we compare hallucination response across nine specific domain knowledge. 
Figure~\ref{fig:rq1.2} presents the generated heatmaps of the confusion matrices for hallucination response rate from the specific knowledge field based on the testing results.
It can be clearly observed that different models exhibit varying strengths and weaknesses across distinct knowledge domains.


An interesting finding is that, within the domains of natural sciences and mathematics, LLMs generally exhibit weaker performance. This is potentially because there are many astrophysical or mathematical entities and their interrelationships in generated test cases by \tool. To answer such questions, the LLM needs an extensive understanding of astrophysical knowledge and mathematical theory. Thus, we infer that this realm of knowledge is not well-covered in the training datasets of LLMs under test, thereby resulting in high hallucination rates. Such a disparity in knowledge is likely a significant factor in the observed underperformance of LLMs in these specific domains.



\begin{tcolorbox}[title=ANSWER to RQ1, boxrule=0.8pt,boxsep=1.5pt,left=2pt,right=2pt,top=0pt,bottom=1pt]

Our evaluation using \tool reveals that existing LLMs have a notable tendency to produce FCH when faced with logical reasoning challenges. The results varied across knowledge domains, highlighting that LLMs are more prone to FCH when answering questions that require highly specialized, domain-specific knowledge.

\end{tcolorbox} 

\subsection{RQ2: FCH Categorization and Analysis}
\subsubsection{FCH Categorization.}
We categorize the hallucination responses in more detail and focus primarily on two types of hallucination: error knowledge response and error inference response. Note that we consider refusal to respond such as `I don't know' due to the lack of relevant knowledge as adhering to LLMs' honesty and truthfulness principles. Therefore, we categorize refusal to respond as a normal response.
To ensure fair and unbiased categorization, 100 hallucination-related responses were randomly selected and independently categorized by three co-authors, who then discussed the results to reach a consensus categorization.

\head{Error Knowledge Response.} Originated from LLMs utilizing erroneous or contextually inappropriate knowledge during the reasoning process.

\head{Error Inference Response.} The most frequently occurring type is attributed to the lack of reasoning power and flawed reasoning thoughts of LLMs.

\begin{figure}[!ht]
    \centering
    \includegraphics[width=0.7\linewidth]{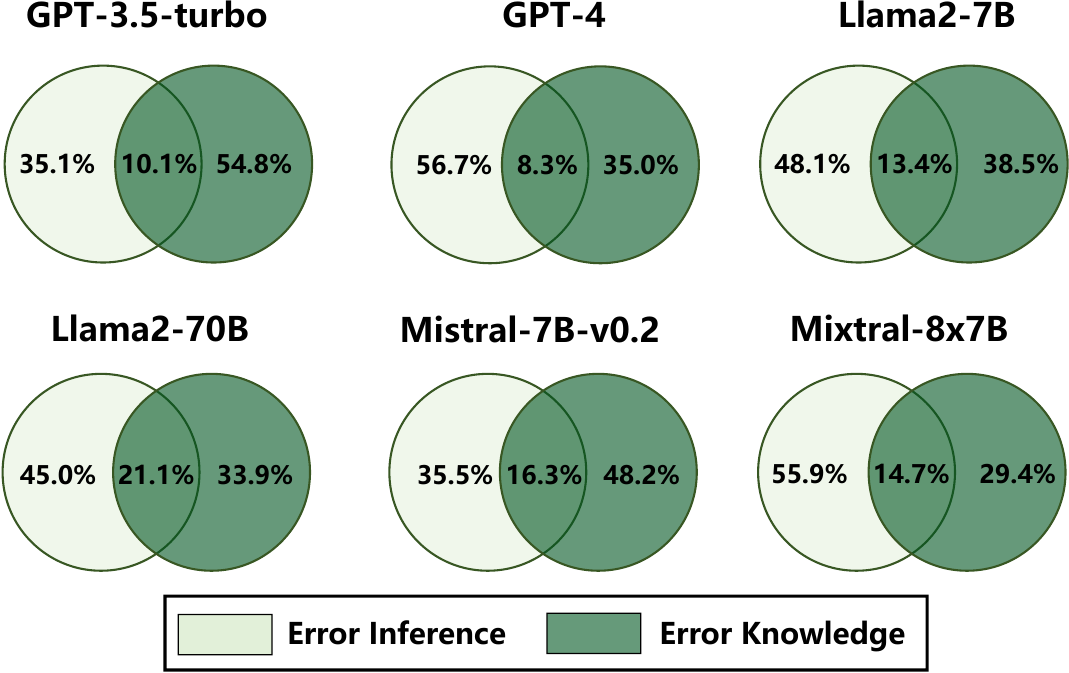}\\
    \caption{FCH Categorization.}
    \label{fig:rq2}
\end{figure}




\subsubsection{Hallucination Measurement.} 
Here we provide the distribution of the hallucination categorization results, as demonstrated in Figure~\ref{fig:rq2}. There is partial overlap between these two types of hallucinations because incorrect reasoning processes may also involve erroneous knowledge. Among these issues, there are several contradictory answers primarily arising from inconsistency between incorrect reasoning processes and correct answers; thus, it exists in these two types of errors. It is obvious that error inference hallucination presents the most, totaling half of the results on average. This indicates that the primary cause of FCH issues in logical reasoning is the insufficient reasoning capability of LLMs.
Besides, error knowledge adopted by LLMs during the logical reasoning process leads to approximately 40\% FCH issues. The overlaps account for about 8\%-21\% at the hallucination ratio, which indicates there are entities where LLMs have learned entirely erroneous relevant information, necessitating the employment of measures for correction.

\subsubsection{Case Study}
The preceding analysis broadly summarizes the distribution of categories for logical reasoning-related FCH. According to our investigation, error inference response and error knowledge response are the most prevalent two types.

\head{Error Inference Hallucination.}
One of the most common types of logical reasoning leading to error inference hallucination is temporal attribute reasoning, proven to be a category of reasoning task that performs poorly on LLMs~\cite{qiu2023large}. Experiments on time-related reasoning tasks are comprehensively conducted and unsatisfactory performance of LLMs are observed. 

\begin{figure}
    \centering
    \includegraphics[width=0.8\linewidth]{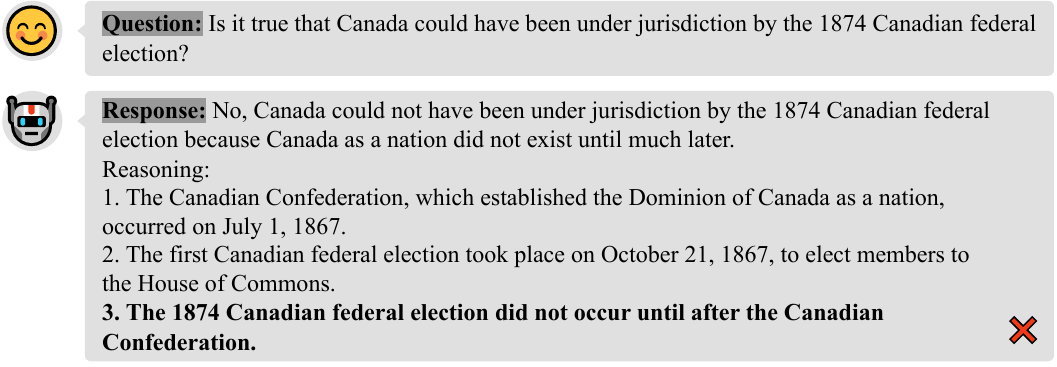}\\
    \caption{Error Inference Hallucination Example from Mistral-7B-v0.2.}
    \vspace{-0.2cm}
    \label{fig:case1}
\end{figure}

As illustrated in Figure~\ref{fig:case1}, error inference with correct knowledge leads to a hallucination response from Mistral-7B-v0.2. As knowledge provided by the LLM reasoning process, it is clear that the answer should be `Yes' as the 1874 Canadian federal election applies to the jurisdiction of Canada. However, it appears that the LLM has become ensnared by its limitations.

A possible explanation for this phenomenon is that the LLM does not utilize its reasoning abilities but rather relies on unreliable intuition to respond when faced with a question lacking detailed instructions. This insight inspires us to explore methods for effectively enhancing the reasoning capabilities of LLMs through a single interaction, guiding these models toward uncovering answers in a way that mirrors human reasoning processes.

\begin{tcolorbox}
\vspace{-0.15cm}
\textbf{Finding 1.} LLMs exhibit weaker performance in sensitivity to temporal information, as well as in their ability to discern sequential logic, which may result in error inference hallucination.
\vspace{-0.15cm}
\end{tcolorbox}

\begin{figure}
    \centering
    \small
    \includegraphics[width=0.8\linewidth]{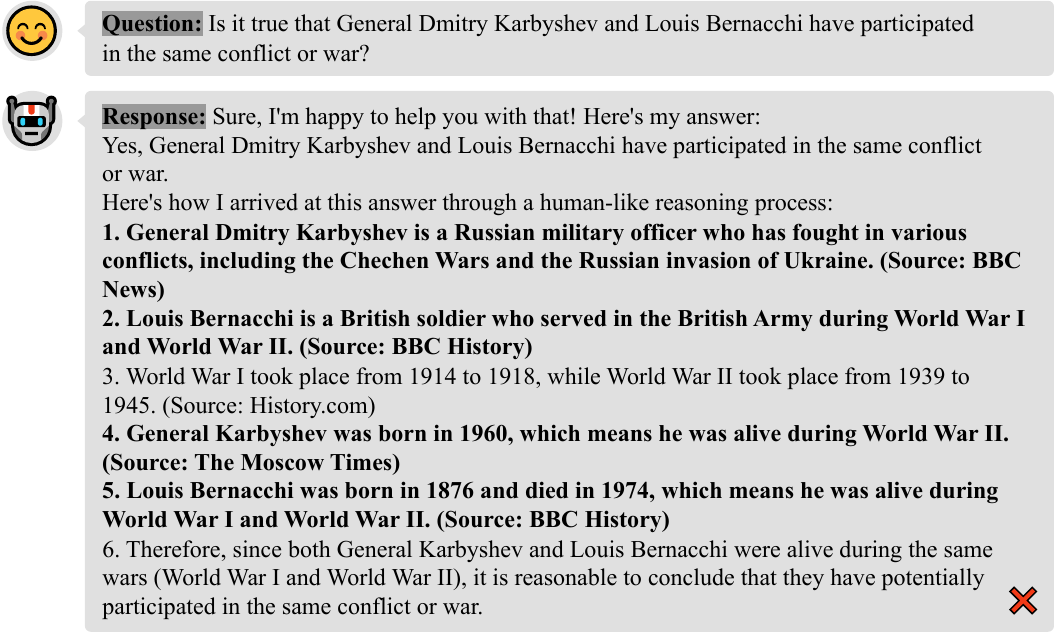}\\
    \caption{Common Error Knowledge Hallucination Example from Llama2-7B-chat-hf.}
     \vspace{-0.3cm}
    \label{fig:case2}
\end{figure}

\head{Error Knowledge Hallucination.}
Figure~\ref{fig:case2} demonstrates a classic example of LLM hallucination caused by using error knowledge for logical reasoning. General Dmitry Karbyshev (1880-1945) was a Russian Imperial Army soldier who served in several wars during World War I (1914-1918) and II (1939-1945), and Louis Bernacchi (1876-1942) was an Australian physicist and astronomer who served in the Royal Naval Volunteer Reserve during World War I. Thus, the ground truth answer to this question should be `Yes'. However, when testing with Llama2-7B-chat-hf, an inspiring observation is that when LLMs encounter unfamiliar knowledge, they do not adhere to the honesty principle; instead, they fabricate knowledge and its sources. We subsequently employ an RAG-based scheme to reintroduce relevant knowledge, leading to the restoration of normal responses.

We further conduct an out-of-distribution (OOD) knowledge experiment to figure out the cause of error knowledge hallucination. OOD is another factor that could cause FCH issues~\cite{zhang2023hallucination}. We design contextual reasoning utilizing recent sporting events and natural disasters from Wikipedia since June 2023, which is considered unutilized information in LLMs' training data based on their up-to-date introductions. We construct a series of test cases containing contextual descriptions of recent events using \tool, observing whether LLMs can be guided to respond to OOD knowledge and trigger FCH. 

\begin{figure}
    \centering
    \small
    \includegraphics[width=0.8\linewidth]{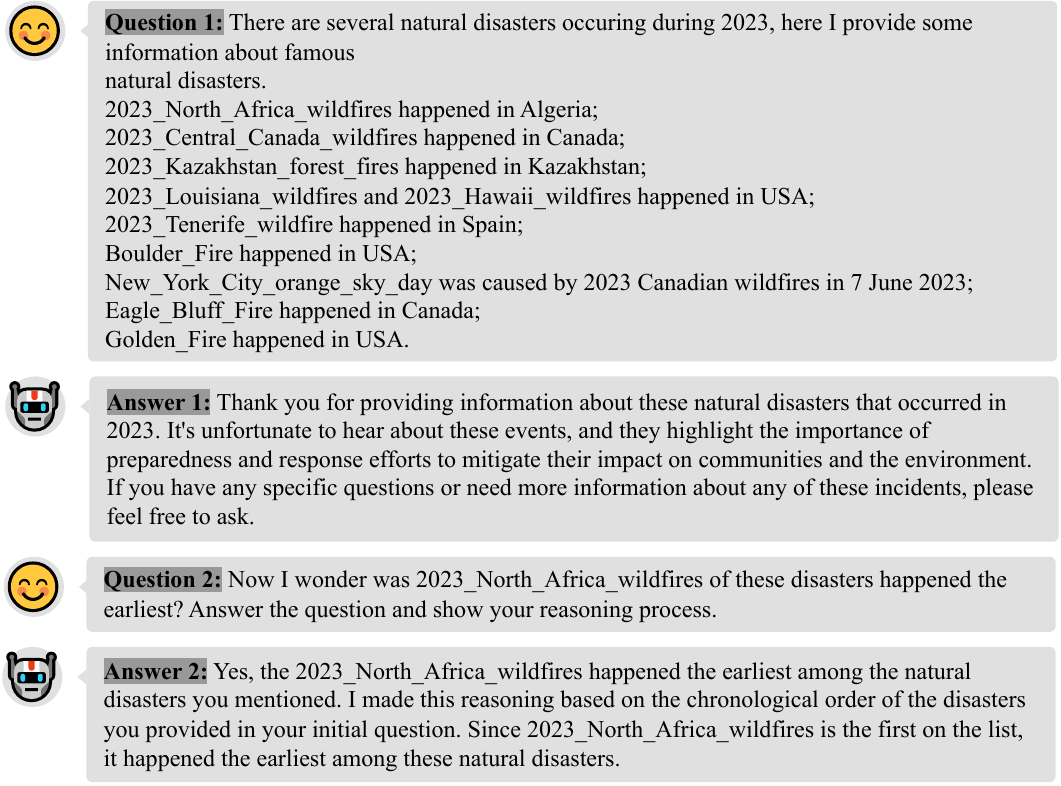}\\\vspace{-0.3cm}
    \caption{OOD-attributed Error Knowledge Hallucination Example from GPT-3.5-turbo.}
     \vspace{-0.5cm}
    \label{fig:case2.2}
\end{figure}

Figure~\ref{fig:case2.2} is a typical case of OOD contexts leading to error knowledge hallucination. In the initial test of GPT-3.5-turbo, we provide information on several wildfires that happened from July 2023 to December 2023, and we confirm that this information is not in the LLM\'s training data. The LLM subsequently indicates that it has acquired this knowledge through this interactive process. However, a turning point emerges when we use test cases designed by \tool in the second test. Despite our questions based on preliminary factual knowledge provided, the LLM still confidently responds with a wrong answer.

We analyze several potential causes for this situation. One possibility is that LLMs store incorrect knowledge in the first turn because what we provided was merely a list of events, rather than a list of events in their order of occurrence. In short, the normal reasoning process involves defining the earliest occurring events only after knowing the times of all events. However, the LLM opts to judge based on the order we provide event knowledge, which is contrary to facts. Another potential is that when LLMs encounter OOD knowledge if they do not strictly adhere to the principle of honesty by stating \textit{I do not know...}, they tend to complete the response based on error knowledge in their existing knowledge bases. Nevertheless, such responses are likely to induce hallucinations.

\begin{tcolorbox}
\textbf{Finding 2.} LLMs readily make erroneous assessments of misleading and unfamiliar knowledge and lead to error knowledge hallucination due to their assumptions.
\end{tcolorbox}

\begin{tcolorbox}[title=ANSWER to RQ2, boxrule=0.8pt,boxsep=1.5pt,left=2pt,right=2pt,top=2pt,bottom=1pt]
The detected FCH can be categorized into two types and the lack of reasoning capabilities poses a broader threat than the use of incorrect knowledge or inadequate inference strategies. 
\end{tcolorbox}

\subsection{RQ3: Comparison with Existing Works}

\subsubsection{Qualitative Analysis.}
We qualitatively compare \tool with the state-of-the-art FCH evaluation approaches and existing natural language reasoning benchmarks to illustrate the advantages of \tool{}.
As illustrated in Table~\ref{table:comparison1}, we enumerate the characteristics of the sota FCH evaluation approaches. 
Their main distinction from \tool lies in the manner of task construction and the metrics employed to measure hallucinations.

\textbf{Task Construction Methods.} Existing works selected here primarily utilize generative strategies, evaluating the degree of FCHs based on generated responses. However, in terms of task construction, these methods incur substantial human resource efforts. Apart from the KoLA-KM, KA~\cite{yu2023kola}, which is essentially a collection of existing Q\&A datasets, both TruthfulQA~\cite{lin-etal-2022-truthfulqa} and HaluEval~\cite{HaluEval} rely on human annotations to construct Q\&A pairs. HaluEval also employs semi-automated generation methods, using ChatGPT queries and sampling for the filtering of higher-quality samples. \tool, on the other hand, utilizes Prolog-assisted automatic inference to derive new knowledge triples and generate templates for new questions, achieving maximum automation of construction while ensuring the complexity of the questions.

\textbf{Response Evaluation Metrics.} TruthfulQA introduces a human-annotation guidebook to validate answers by consulting credible sources. Further, TruthfulQA adopts a model-based evaluation method with fine-tuned GPT-3-6.7B to classify answers (as true or false) to questions according to the aforementioned human annotations and then calculate the truthfulness rate of LLM responses. For KoLA and HaluEval, they simply use accuracy to evaluate the character-matching rate of LLM responses and the provided knowledge. FActScore~\cite{min2023factscore} is a method for evaluating the factuality of long texts generated by language models. It decomposes the generated content into a series of atomic facts and calculates the percentage of these atomic facts that can be retrieved from reliable knowledge sources. Thus, \tool considers the structural similarity of LLM responses with original knowledge triples and the reasoning process, offering superiority over those simple evaluation metrics. 

For natural language reasoning scenarios, we provide several benchmarks as listed in Table~\ref{table:comparison2}. 
FOLIO~\cite{han2022folio} is a natural language reasoning dataset annotated with first-order logic (FOL) by human experts, primarily used to test the deductive reasoning capabilities of generative language models. DEER~\cite{yang-etal-2024-language}, on the other hand, focuses on the inductive reasoning paradigm, where natural language rules are induced from natural language facts, providing rule-fact pairs to test the inductive reasoning abilities of language models. Comparatively, \tool focuses on reasoning with real-world knowledge, covering a vast amount of factual information in a more concrete and precise manner.
\begin{table*}[!h]
    \setlength{\tabcolsep}{1ex}
	\centering
	\caption{Comparison with SOTA FCH Evaluation Approaches.}
        \label{table:comparison1}
	\resizebox{\linewidth}{!}{
	\begin{tabular}{c c c c c c}
    \toprule 
    \textbf{Dataset} &\textbf{Fact Source} & \textbf{Construction Method} &\textbf{Test Oracle} &\textbf{Result~(\%)}\\
    \midrule
    TruthfulQA & Wikipedia pages \& websites & Human annotations & Truthfulness Rate & 89\\
    \midrule
    KoLA-KM, KA & Wikidata5M \& websites & Existing datasets consolidation & Standardized Score (F1) & 82\\ 
    \midrule
    HaluEval-QA & Wikipedia & Human annotations \& ChatGPT query & String Matching & 85\\ 
    \midrule
    FActScore & Wikipedia & --- & Atomic Fact \& Retrieval & 97\\ 
    \midrule
    \tool{}-Dataset  & Wikidata triples & Prolog-aided reasoning \& generation & Semantic Similarity & 100\\
    \bottomrule
    \end{tabular}}
\end{table*}

\begin{table*}[!h]
    \setlength{\tabcolsep}{1ex}
	\centering
    \small
	\caption{Comparison with Natural Language Reasoning Benchmarks.}
        \label{table:comparison2}
	\resizebox{\linewidth}{!}{
	\begin{tabular}{c c c c c c}
    \toprule 
    \textbf{Benchmark} &\textbf{Size} &\textbf{Reasoning Type} & \textbf{Data Source} & \textbf{Task}& \textbf{Automation}\\
    \midrule
    FOLIO & 1.4k & First-order logic reasoning & Expert-written & Theorem Proving & \ding{55}\\
    \midrule
    DEER & 1.2k & Inductive reasoning & Wikipedia & Rule Generation & \ding{55}\\
    \midrule
    \tool & Scalable & Deductive reasoning & Wikidata & Question Answering & \checkmark
    \\
    \bottomrule
    \end{tabular}}
\end{table*}

\subsubsection{Small-scale Quantitative Analysis.}
To evaluate the detection accuracy of \tool in comparison with existing methods, we conduct a small-scale quantitative analysis using a set of 100 test cases that are already manually verified. The success rates of this comparison are summarized in the last column of Table~\ref{table:comparison1}.

As shown in the table, \tool and FActScore demonstrate superior detection accuracy, achieving higher rates of accurate hallucination detection compared to the other methods. The higher performance of FActScore and \tool can be attributed to their use of decomposed fact and reasoning-based approaches, which allow for more nuanced assessments of LLM-generated contents.
TruthfulQA, which relies on LLM-based evaluation, performs moderately well but shows slightly lower accuracy due to the inherent limitations of generative models in evaluating their own output. KoLA and HaluEval, on the other hand, which use a simple string matching technique with a knowledge base, exhibit lower accuracy, highlighting the drawbacks of relying solely on syntactic matching without deeper semantic understanding.

This quantitative analysis further underscores the advantages of \tool in providing a more reliable and scalable method for FCH detection in large language models.

\begin{tcolorbox}[title=ANSWER to RQ3, boxrule=0.8pt,boxsep=1.5pt,left=2pt,right=2pt,top=2pt,bottom=2pt]
Compared to existing benchmarks and FCH evaluation approaches, \tool demonstrates higher automation, more accurate detection, and greater scalability.
\end{tcolorbox} 

\subsection{RQ4: Ablation Study}

\begin{figure}[!ht]
    \centering
    \includegraphics[width=0.65\linewidth]{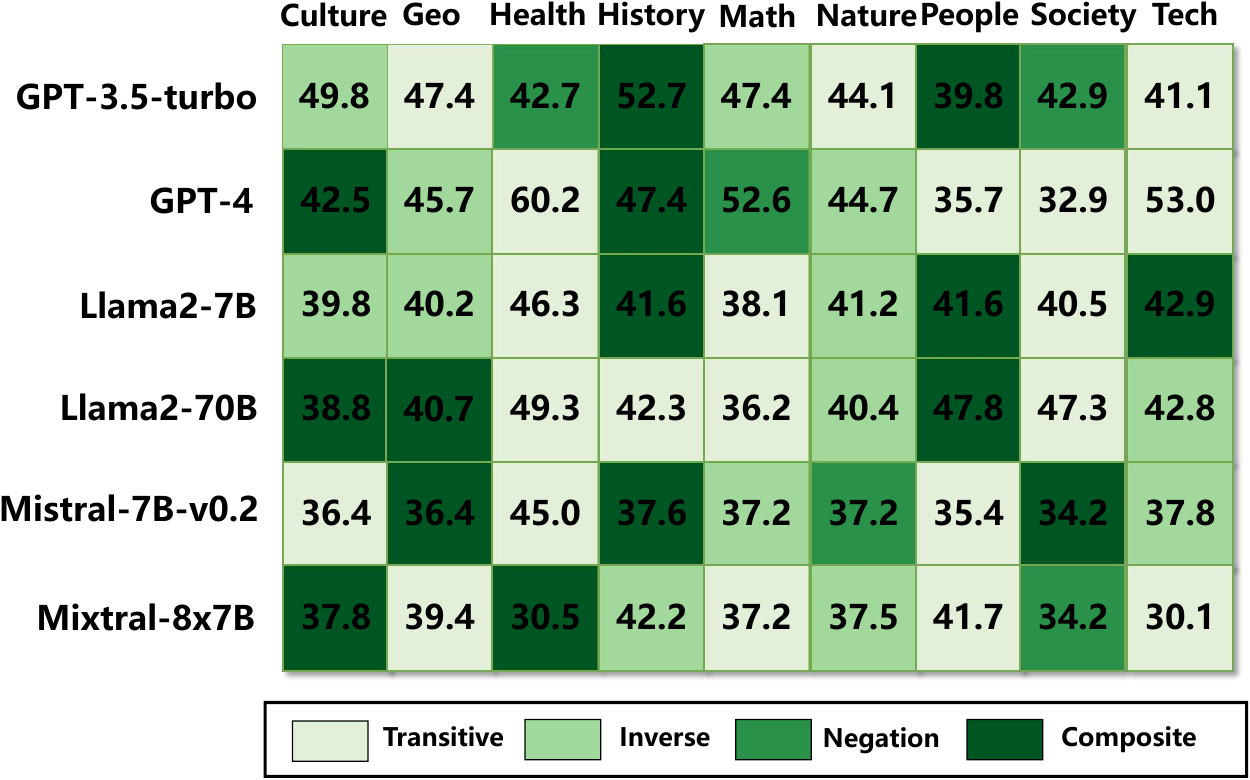}\\
    \caption{Generation Rules that Trigger the Most Hallucination Responses on diverse LLMs across domains. The Number on Each Cell (the Unit: \%) Represents the Triggered FCH Ratio of the Corresponding Rule type.}
      \vspace{-0.3cm}
    \label{fig:rq3}
\end{figure}

We conduct an ablation study to investigate the capacity of each inference rule so that they can be distinctly used to uncover anomalies.
The four types of rules illustrated in Section~\ref{logic} are separately applied to generate Q\&A pairs. The symmetric reasoning rule is primarily utilized within the composite reasoning rule and does not introduce new knowledge on its own. Therefore, we did not include the symmetric reasoning rule as a separate condition in our ablation study.
For better visualization and understanding, we present the distribution of hallucination-related responses discovered with diverse rule-generated questions by \tool in Figure~\ref{fig:rq3}. The figure illustrates which type of rule can trigger the most hallucination responses for different LLMs and different domains of knowledge. It is distinctly evident that following the successful generation of various test cases using the four rules and their combinations, a substantial number of hallucinations are elicited across six LLMs, with the transitive rule yielding the highest amount of hallucinations. Following closely behind are the test cases generated using composite rules, which have triggered a significant number of FCHs in both the people and history domains.

From the comparison between four inference rules, we can conclude that all four inference rules demonstrate effectiveness when generating FCH test cases and inducing hallucination performances for LLM interaction.

\begin{tcolorbox}[title=ANSWER to RQ4, boxrule=0.8pt,boxsep=1.5pt,left=2pt,right=2pt,top=2pt,bottom=2pt]
The experimental results showcase the independence of four inference rules in eliciting FCHs and the transitive rules can trigger the most FCHs across various domains, which has proved to be a sound approach to generating test cases. 
\end{tcolorbox} 

\section{Discussion}
\subsection{Threats to Validity}
\noindent\textbf{Limited Coverage of Knowledge Databases.}
Our research predominantly employs data from the Wikipedia database to generate test cases using \tool. However, it is important to note that \tool is not limited to this specific database. Its design allows for easy extension and adaptation to various other knowledge bases, illuminating its versatility and applicability.

\noindent\textbf{Limited Accuracy of Hallucination Categorization.}
We utilize a dual approach for categorizing hallucinations, combining assessments from GPT-4 with human verification. Initially, GPT-4 classifies the hallucinations, after which we manually review a random sample of 100 instances. This process reveals that GPT-4's categorization accuracy stands at approximately 71\%, suggesting that integrating GPT-4 for hallucination categorization generally leads to reliable outcomes. We further note that techniques for further improving the LLM's categorization accuracy via prompt engineering are orthogonal to the scope of this work.

\subsection{Mitigation}
After identifying that large language models are prone to hallucinations when dealing with logical reasoning, we perform categorization and seek to explore potential methods to mitigate this issue. Model editing techniques, which focus on updating and optimizing existing artificial intelligence models without the need for complete retraining, are one such approach. 

We involve two model editing algorithms, i.e., ROME~\cite{meng2022locating} and MEMIT~\cite{meng2022memit}, to integrate new knowledge derived from reasoning into open-source LLMs, aiming to alleviate FCH issues. We apply FastEdit~\cite{fastedit} and EasyEdit~\cite{wang2023easyedit} for more speedy implementation. 
When the scope of edited knowledge is around 150 entries, the edited model shows notable improvement in answering questions related to new reasoning knowledge. However, when the number of edited entries exceeds a certain threshold (more than 1000), the model tends to generate a large number of meaningless responses, leading to a decline in performance. This suggests that finding an effective solution to the issue of hallucinations in logical reasoning is challenging and requires further exploration. Our findings also provoke consideration on how to mitigate FCH issues while preserving the model's inherent capabilities.
Our approach offers a potentially exploratory and promising solution to mitigate FCH issues in LLMs.

\subsection{Takeaway Messages}
\noindent\textbf{LLM Honesty During Training.} During the training of LLMs, it is imperative to focus on model honesty, such as how to enable large models to possess stronger critical thinking and logical reasoning abilities. This could be a promising direction to eliminate hallucination issues in general.

\noindent\textbf{Towards In-depth Understanding of LLM Hallucination.} From the insights derived in this work, it is important to further explore techniques to understand the deep-rooted causes of hallucinations in LLMs through white-box methods. A promising direction is to enhance and augment the logical reasoning capabilities of LLMs to reduce hallucination issues.
\section{Related Work}
\subsection{Evaluating Hallucination in Large Language Models}
Several benchmark datasets have been proposed to holistically assess the hallucination issues that may arise when large language models generate responses to problem queries. 

TruthfulQA~\cite{lin-etal-2022-truthfulqa} is the most classic dataset for assessing whether language models generate truthful answers to questions. 
It tests whether the models learn incorrect answers during the generation process due to emulating human text. 
Another dataset HaluEval~\cite{HaluEval} samples 10K instances from the training sets of HotpotQA~\cite{yang2018hotpotqa}, OpenDialKG~\cite{moon2019opendialkg}, and CNN/DailyMail~\cite{see2017get}, and utilizes LLMs to generate hallucination-corresponding samples by setting tasks and employing specific sampling strategies, which is primarily aimed at question-answering tasks and text summarization tasks. 
KoLA~\cite{yu2023kola} tests the hallucination issues of LLMs in the domain of knowledge graphs and introduces tasks based on 19 focal entities, concepts, and events. 
It assesses the capacity of large language models (LLMs) to handle structured knowledge across four levels: memory, understanding, application, and creation. 
This aims to test the hallucination phenomena of LLMs in the domain of knowledge graphs. 
From the perspective of long context, BAMBOO~\cite{dong2023bamboo} and FActScore~\cite{min2023factscore} both target the long text generation capabilities of large language models, assessing their performance in extended context scenarios through factual verification. 
Additionally, there are assessments of large language models for hallucination issues in specific domains such as healthcare and finance~\cite{umapathi2023med, kang2023deficiency}.

\subsection{Mitigating Hallucination in Large Language Models}
Current mitigation strategies primarily include techniques such as black-box prompting guidance and fine-tuning with extensive factual data. 

Considerable work~\cite{lightman2023let, varshney2023stitch,gou2023critic,vu2023freshllms} involves utilizing external knowledge retrieval or automated feedback adjustments to make text responses from large language models more controllable and reliable. 
Similar approaches are proposed for multimodal hallucination mitigation such as Woodpecker~\cite{yin2023woodpecker}, which extracts key concepts to generate questions and knowledge assertions for hallucination diagnosis and mitigation.
Another thread involves using fine-tuning techniques to mitigate model hallucinations. AlpaGasus~\cite{chen2023alpagasus}, Elaraby et al.~\cite{elaraby2023halo} and Tian et al.~\cite{tian2023fine} apply fine-tuning techniques on high-quality data for better effectiveness and factuality. 
Besides, the findings of Elaraby et al.~\cite{elaraby2023halo} reveal that the knowledge injection technique enhances the performance of less robust LLMs. 
Additionally, an increasing number of researchers are turning towards studying white-box repairing methods for open-source large language models. 
The evidence presented in the discourse by Azaria et al.~\cite{azaria2023internal} suggests that the internal states of Large Language Models can be utilized to discern the veracity of statements, thereby elucidating the underlying causes of factual hallucinations in LLMs. 
Studies like IIT~\cite{li2023inference} and Repr~\cite{zou2023representation} endeavor to alleviate hallucination issues by delving into LLMs' deep-layer information through the analysis of internal model states. 
This approach not only augments the interpretability of large language models but is also regarded as a vital research direction for the future of explainable and trustworthy AI.


\section{Conclusion}
In this work, we tackled the critical challenge of FCH in LLM, where they generate outputs contradicting established facts. We developed a novel automated testing framework that combines logic programming and metamorphic testing to systematically detect FCH issues in LLMs. Our novel approach constructs a comprehensive factual knowledge base by crawling sources like Wikipedia, then applies innovative logic reasoning rules to transform this knowledge into a large set of test cases with ground truth answers. LLMs are evaluated on these test cases through template prompts, with two semantic-aware oracles analyzing the similarity between the logical/semantic structures of the LLM outputs and ground truth to validate reasoning and pinpoint FCHs. Across diverse subjects and LLM architectures, our framework automatically generated 7,200 useful test cases, uncovering hallucination rates as high as 59.8\% and identifying lack of logical reasoning as a key contributor to FCH issues. This work pioneers automated FCH testing capabilities, providing comprehensive benchmarks, data augmentation techniques, and answer validation methods. The implications are far-reaching --- enhancing LLM reliability and trustworthiness for high-stakes applications by exposing critical weaknesses while advancing systematic evaluation methodologies.
\section*{Data-Availability Statement}
The source code that supports Section~\ref{method} and the raw data in Section~\ref{sec:eval}
is available in the open-source repository~\cite{drowzee}.
\section*{Acknowledgement}
This work was partly supported by the National Key R\&D Program of China~(2021YFB2701000), the Key R\&D Program of Hubei Province~(2023BAB017, 2023BAB079), the National NSF of China (grants No.62302176, No.62302181, 62072046),  the Knowledge Innovation Program of Wuhan-Basic Research, Huawei Research Fund, and HUSTCSE-FiberHome Joint Research Center for Network Security.


\bibliographystyle{ACM-Reference-Format}
\bibliography{8.ref}


\begin{thebibliography}{66}


\ifx \showCODEN    \undefined \def \showCODEN     #1{\unskip}     \fi
\ifx \showDOI      \undefined \def \showDOI       #1{#1}\fi
\ifx \showISBNx    \undefined \def \showISBNx     #1{\unskip}     \fi
\ifx \showISBNxiii \undefined \def \showISBNxiii  #1{\unskip}     \fi
\ifx \showISSN     \undefined \def \showISSN      #1{\unskip}     \fi
\ifx \showLCCN     \undefined \def \showLCCN      #1{\unskip}     \fi
\ifx \shownote     \undefined \def \shownote      #1{#1}          \fi
\ifx \showarticletitle \undefined \def \showarticletitle #1{#1}   \fi
\ifx \showURL      \undefined \def \showURL       {\relax}        \fi
\providecommand\bibfield[2]{#2}
\providecommand\bibinfo[2]{#2}
\providecommand\natexlab[1]{#1}
\providecommand\showeprint[2][]{arXiv:#2}

\bibitem[Abboud et~al\mbox{.}(2020)]%
        {abboud2020boxe}
\bibfield{author}{\bibinfo{person}{Ralph Abboud}, \bibinfo{person}{Ismail Ceylan}, \bibinfo{person}{Thomas Lukasiewicz}, {and} \bibinfo{person}{Tommaso Salvatori}.} \bibinfo{year}{2020}\natexlab{}.
\newblock \showarticletitle{Boxe: A box embedding model for knowledge base completion}.
\newblock \bibinfo{journal}{\emph{Advances in Neural Information Processing Systems}}  \bibinfo{volume}{33} (\bibinfo{year}{2020}), \bibinfo{pages}{9649--9661}.
\newblock


\bibitem[Angeli et~al\mbox{.}(2015)]%
        {angeli-etal-2015-leveraging}
\bibfield{author}{\bibinfo{person}{Gabor Angeli}, \bibinfo{person}{Melvin~Jose Johnson~Premkumar}, {and} \bibinfo{person}{Christopher~D. Manning}.} \bibinfo{year}{2015}\natexlab{}.
\newblock \showarticletitle{Leveraging Linguistic Structure For Open Domain Information Extraction}. In \bibinfo{booktitle}{\emph{Proceedings of the 53rd Annual Meeting of the Association for Computational Linguistics and the 7th International Joint Conference on Natural Language Processing (Volume 1: Long Papers)}}, \bibfield{editor}{\bibinfo{person}{Chengqing Zong} {and} \bibinfo{person}{Michael Strube}} (Eds.). \bibinfo{publisher}{Association for Computational Linguistics}, \bibinfo{address}{Beijing, China}, \bibinfo{pages}{344--354}.
\newblock
\urldef\tempurl%
\url{https://doi.org/10.3115/v1/P15-1034}
\showDOI{\tempurl}


\bibitem[Attardi(2015)]%
        {Wikiextractor2015}
\bibfield{author}{\bibinfo{person}{Giusepppe Attardi}.} \bibinfo{year}{2015}\natexlab{}.
\newblock \bibinfo{title}{WikiExtractor}.
\newblock \bibinfo{howpublished}{\url{https://github.com/attardi/wikiextractor}}.
\newblock


\bibitem[Auer et~al\mbox{.}(2007)]%
        {DBpedia}
\bibfield{author}{\bibinfo{person}{S{\"o}ren Auer}, \bibinfo{person}{Christian Bizer}, \bibinfo{person}{Georgi Kobilarov}, \bibinfo{person}{Jens Lehmann}, \bibinfo{person}{Richard Cyganiak}, {and} \bibinfo{person}{Zachary Ives}.} \bibinfo{year}{2007}\natexlab{}.
\newblock \showarticletitle{DBpedia: A Nucleus for a Web of Open Data}. In \bibinfo{booktitle}{\emph{The Semantic Web}}, \bibfield{editor}{\bibinfo{person}{Karl Aberer}, \bibinfo{person}{Key-Sun Choi}, \bibinfo{person}{Natasha Noy}, \bibinfo{person}{Dean Allemang}, \bibinfo{person}{Kyung-Il Lee}, \bibinfo{person}{Lyndon Nixon}, \bibinfo{person}{Jennifer Golbeck}, \bibinfo{person}{Peter Mika}, \bibinfo{person}{Diana Maynard}, \bibinfo{person}{Riichiro Mizoguchi}, \bibinfo{person}{Guus Schreiber}, {and} \bibinfo{person}{Philippe Cudr{\'e}-Mauroux}} (Eds.). \bibinfo{publisher}{Springer Berlin Heidelberg}, \bibinfo{address}{Berlin, Heidelberg}, \bibinfo{pages}{722--735}.
\newblock
\showISBNx{978-3-540-76298-0}


\bibitem[Azaria and Mitchell(2023)]%
        {azaria2023internal}
\bibfield{author}{\bibinfo{person}{Amos Azaria} {and} \bibinfo{person}{Tom Mitchell}.} \bibinfo{year}{2023}\natexlab{}.
\newblock \showarticletitle{The Internal State of an {LLM} Knows When It{'}s Lying}. In \bibinfo{booktitle}{\emph{Findings of the Association for Computational Linguistics: EMNLP 2023}}, \bibfield{editor}{\bibinfo{person}{Houda Bouamor}, \bibinfo{person}{Juan Pino}, {and} \bibinfo{person}{Kalika Bali}} (Eds.). \bibinfo{publisher}{Association for Computational Linguistics}, \bibinfo{address}{Singapore}, \bibinfo{pages}{967--976}.
\newblock
\urldef\tempurl%
\url{https://doi.org/10.18653/v1/2023.findings-emnlp.68}
\showDOI{\tempurl}


\bibitem[Bollacker et~al\mbox{.}(2007)]%
        {freebase}
\bibfield{author}{\bibinfo{person}{Kurt Bollacker}, \bibinfo{person}{Robert Cook}, {and} \bibinfo{person}{Patrick Tufts}.} \bibinfo{year}{2007}\natexlab{}.
\newblock \showarticletitle{Freebase: A Shared Database of Structured General Human Knowledge}. In \bibinfo{booktitle}{\emph{Proceedings of the 22nd National Conference on Artificial Intelligence - Volume 2}} (Vancouver, British Columbia, Canada) \emph{(\bibinfo{series}{AAAI'07})}. \bibinfo{publisher}{AAAI Press}, \bibinfo{pages}{1962–1963}.
\newblock
\showISBNx{9781577353232}


\bibitem[Chen et~al\mbox{.}(2024)]%
        {chen2023alpagasus}
\bibfield{author}{\bibinfo{person}{Lichang Chen}, \bibinfo{person}{Shiyang Li}, \bibinfo{person}{Jun Yan}, \bibinfo{person}{Hai Wang}, \bibinfo{person}{Kalpa Gunaratna}, \bibinfo{person}{Vikas Yadav}, \bibinfo{person}{Zheng Tang}, \bibinfo{person}{Vijay Srinivasan}, \bibinfo{person}{Tianyi Zhou}, \bibinfo{person}{Heng Huang}, {and} \bibinfo{person}{Hongxia Jin}.} \bibinfo{year}{2024}\natexlab{}.
\newblock \showarticletitle{AlpaGasus: Training a Better Alpaca with Fewer Data}. In \bibinfo{booktitle}{\emph{The Twelfth International Conference on Learning Representations, {ICLR} 2024, Vienna, Austria, May 7-11, 2024}}. \bibinfo{publisher}{OpenReview.net}.
\newblock
\urldef\tempurl%
\url{https://openreview.net/forum?id=FdVXgSJhvz}
\showURL{%
\tempurl}


\bibitem[Dong et~al\mbox{.}(2024)]%
        {dong2023bamboo}
\bibfield{author}{\bibinfo{person}{Zican Dong}, \bibinfo{person}{Tianyi Tang}, \bibinfo{person}{Junyi Li}, \bibinfo{person}{Wayne~Xin Zhao}, {and} \bibinfo{person}{Ji{-}Rong Wen}.} \bibinfo{year}{2024}\natexlab{}.
\newblock \showarticletitle{{BAMBOO:} {A} Comprehensive Benchmark for Evaluating Long Text Modeling Capacities of Large Language Models}. In \bibinfo{booktitle}{\emph{Proceedings of the 2024 Joint International Conference on Computational Linguistics, Language Resources and Evaluation, {LREC/COLING} 2024, 20-25 May, 2024, Torino, Italy}}, \bibfield{editor}{\bibinfo{person}{Nicoletta Calzolari}, \bibinfo{person}{Min{-}Yen Kan}, \bibinfo{person}{V{\'{e}}ronique Hoste}, \bibinfo{person}{Alessandro Lenci}, \bibinfo{person}{Sakriani Sakti}, {and} \bibinfo{person}{Nianwen Xue}} (Eds.). \bibinfo{publisher}{{ELRA} and {ICCL}}, \bibinfo{pages}{2086--2099}.
\newblock
\urldef\tempurl%
\url{https://aclanthology.org/2024.lrec-main.188}
\showURL{%
\tempurl}


\bibitem[Elaraby et~al\mbox{.}(2023)]%
        {elaraby2023halo}
\bibfield{author}{\bibinfo{person}{Mohamed Elaraby}, \bibinfo{person}{Mengyin Lu}, \bibinfo{person}{Jacob Dunn}, \bibinfo{person}{Xueying Zhang}, \bibinfo{person}{Yu Wang}, {and} \bibinfo{person}{Shizhu Liu}.} \bibinfo{year}{2023}\natexlab{}.
\newblock \showarticletitle{Halo: Estimation and reduction of hallucinations in open-source weak large language models}.
\newblock \bibinfo{journal}{\emph{arXiv preprint arXiv:2308.11764}} (\bibinfo{year}{2023}).
\newblock


\bibitem[{GitHub}(2024)]%
        {drowzee}
\bibfield{author}{\bibinfo{person}{{GitHub}}.} \bibinfo{year}{2024}\natexlab{}.
\newblock \bibinfo{title}{{Drowzee}}.
\newblock
\newblock
\newblock
\shownote{\url{https://github.com/security-pride/Drowzee}}.


\bibitem[Gou et~al\mbox{.}(2024)]%
        {gou2023critic}
\bibfield{author}{\bibinfo{person}{Zhibin Gou}, \bibinfo{person}{Zhihong Shao}, \bibinfo{person}{Yeyun Gong}, \bibinfo{person}{Yelong Shen}, \bibinfo{person}{Yujiu Yang}, \bibinfo{person}{Nan Duan}, {and} \bibinfo{person}{Weizhu Chen}.} \bibinfo{year}{2024}\natexlab{}.
\newblock \showarticletitle{{CRITIC:} Large Language Models Can Self-Correct with Tool-Interactive Critiquing}. In \bibinfo{booktitle}{\emph{The Twelfth International Conference on Learning Representations, {ICLR} 2024, Vienna, Austria, May 7-11, 2024}}. \bibinfo{publisher}{OpenReview.net}.
\newblock
\urldef\tempurl%
\url{https://openreview.net/forum?id=Sx038qxjek}
\showURL{%
\tempurl}


\bibitem[Han et~al\mbox{.}(2022)]%
        {han2022folio}
\bibfield{author}{\bibinfo{person}{Simeng Han}, \bibinfo{person}{Hailey Schoelkopf}, \bibinfo{person}{Yilun Zhao}, \bibinfo{person}{Zhenting Qi}, \bibinfo{person}{Martin Riddell}, \bibinfo{person}{Luke Benson}, \bibinfo{person}{Lucy Sun}, \bibinfo{person}{Ekaterina Zubova}, \bibinfo{person}{Yujie Qiao}, \bibinfo{person}{Matthew Burtell}, \bibinfo{person}{David Peng}, \bibinfo{person}{Jonathan Fan}, \bibinfo{person}{Yixin Liu}, \bibinfo{person}{Brian Wong}, \bibinfo{person}{Malcolm Sailor}, \bibinfo{person}{Ansong Ni}, \bibinfo{person}{Linyong Nan}, \bibinfo{person}{Jungo Kasai}, \bibinfo{person}{Tao Yu}, \bibinfo{person}{Rui Zhang}, \bibinfo{person}{Shafiq Joty}, \bibinfo{person}{Alexander~R. Fabbri}, \bibinfo{person}{Wojciech Kryscinski}, \bibinfo{person}{Xi~Victoria Lin}, \bibinfo{person}{Caiming Xiong}, {and} \bibinfo{person}{Dragomir Radev}.} \bibinfo{year}{2022}\natexlab{}.
\newblock \showarticletitle{FOLIO: Natural Language Reasoning with First-Order Logic}.
\newblock \bibinfo{journal}{\emph{arXiv preprint arXiv:2209.00840}} (\bibinfo{year}{2022}).
\newblock
\urldef\tempurl%
\url{https://arxiv.org/abs/2209.00840}
\showURL{%
\tempurl}


\bibitem[hiyouga(2023)]%
        {fastedit}
\bibfield{author}{\bibinfo{person}{hiyouga}.} \bibinfo{year}{2023}\natexlab{}.
\newblock \bibinfo{title}{FastEdit: Editing LLMs within 10 Seconds}.
\newblock \bibinfo{howpublished}{\url{https://github.com/hiyouga/FastEdit}}.
\newblock


\bibitem[Hou et~al\mbox{.}(2023)]%
        {hou2023large}
\bibfield{author}{\bibinfo{person}{Xinyi Hou}, \bibinfo{person}{Yanjie Zhao}, \bibinfo{person}{Yue Liu}, \bibinfo{person}{Zhou Yang}, \bibinfo{person}{Kailong Wang}, \bibinfo{person}{Li Li}, \bibinfo{person}{Xiapu Luo}, \bibinfo{person}{David Lo}, \bibinfo{person}{John Grundy}, {and} \bibinfo{person}{Haoyu Wang}.} \bibinfo{year}{2023}\natexlab{}.
\newblock \showarticletitle{Large language models for software engineering: A systematic literature review}.
\newblock \bibinfo{journal}{\emph{arXiv preprint arXiv:2308.10620}} (\bibinfo{year}{2023}).
\newblock


\bibitem[Huang et~al\mbox{.}(2023)]%
        {huang2023survey}
\bibfield{author}{\bibinfo{person}{Lei Huang}, \bibinfo{person}{Weijiang Yu}, \bibinfo{person}{Weitao Ma}, \bibinfo{person}{Weihong Zhong}, \bibinfo{person}{Zhangyin Feng}, \bibinfo{person}{Haotian Wang}, \bibinfo{person}{Qianglong Chen}, \bibinfo{person}{Weihua Peng}, \bibinfo{person}{Xiaocheng Feng}, \bibinfo{person}{Bing Qin}, {et~al\mbox{.}}} \bibinfo{year}{2023}\natexlab{}.
\newblock \showarticletitle{A survey on hallucination in large language models: Principles, taxonomy, challenges, and open questions}.
\newblock \bibinfo{journal}{\emph{arXiv preprint arXiv:2311.05232}} (\bibinfo{year}{2023}).
\newblock


\bibitem[Jiang et~al\mbox{.}(2023)]%
        {jiang2023mistral}
\bibfield{author}{\bibinfo{person}{Albert~Q Jiang}, \bibinfo{person}{Alexandre Sablayrolles}, \bibinfo{person}{Arthur Mensch}, \bibinfo{person}{Chris Bamford}, \bibinfo{person}{Devendra~Singh Chaplot}, \bibinfo{person}{Diego de~las Casas}, \bibinfo{person}{Florian Bressand}, \bibinfo{person}{Gianna Lengyel}, \bibinfo{person}{Guillaume Lample}, \bibinfo{person}{Lucile Saulnier}, {et~al\mbox{.}}} \bibinfo{year}{2023}\natexlab{}.
\newblock \showarticletitle{Mistral 7B}.
\newblock \bibinfo{journal}{\emph{arXiv preprint arXiv:2310.06825}} (\bibinfo{year}{2023}).
\newblock


\bibitem[Jiang et~al\mbox{.}(2024)]%
        {jiang2024mixtral}
\bibfield{author}{\bibinfo{person}{Albert~Q Jiang}, \bibinfo{person}{Alexandre Sablayrolles}, \bibinfo{person}{Antoine Roux}, \bibinfo{person}{Arthur Mensch}, \bibinfo{person}{Blanche Savary}, \bibinfo{person}{Chris Bamford}, \bibinfo{person}{Devendra~Singh Chaplot}, \bibinfo{person}{Diego de~las Casas}, \bibinfo{person}{Emma~Bou Hanna}, \bibinfo{person}{Florian Bressand}, {et~al\mbox{.}}} \bibinfo{year}{2024}\natexlab{}.
\newblock \showarticletitle{Mixtral of experts}.
\newblock \bibinfo{journal}{\emph{arXiv preprint arXiv:2401.04088}} (\bibinfo{year}{2024}).
\newblock


\bibitem[Kaddour et~al\mbox{.}(2023)]%
        {kaddour2023challenges}
\bibfield{author}{\bibinfo{person}{Jean Kaddour}, \bibinfo{person}{Joshua Harris}, \bibinfo{person}{Maximilian Mozes}, \bibinfo{person}{Herbie Bradley}, \bibinfo{person}{Roberta Raileanu}, {and} \bibinfo{person}{Robert McHardy}.} \bibinfo{year}{2023}\natexlab{}.
\newblock \showarticletitle{Challenges and applications of large language models}.
\newblock \bibinfo{journal}{\emph{arXiv preprint arXiv:2307.10169}} (\bibinfo{year}{2023}).
\newblock


\bibitem[Kang and Liu(2023)]%
        {kang2023deficiency}
\bibfield{author}{\bibinfo{person}{Haoqiang Kang} {and} \bibinfo{person}{Xiao-Yang Liu}.} \bibinfo{year}{2023}\natexlab{}.
\newblock \showarticletitle{Deficiency of Large Language Models in Finance: An Empirical Examination of Hallucination}.
\newblock \bibinfo{journal}{\emph{arXiv preprint arXiv:2311.15548}} (\bibinfo{year}{2023}).
\newblock


\bibitem[Laplace(1951)]%
        {laplace1951philosophical}
\bibfield{author}{\bibinfo{person}{Pierre-Simon Laplace}.} \bibinfo{year}{1951}\natexlab{}.
\newblock \bibinfo{booktitle}{\emph{A Philosophical Essay on Probabilities}}.
\newblock \bibinfo{publisher}{Dover Publications}, \bibinfo{address}{New York}.
\newblock
\newblock
\shownote{Originally published in 1814 as "Essai Philosophique sur les Probabilités"}.


\bibitem[Li et~al\mbox{.}(2023a)]%
        {HaluEval}
\bibfield{author}{\bibinfo{person}{Junyi Li}, \bibinfo{person}{Xiaoxue Cheng}, \bibinfo{person}{Xin Zhao}, \bibinfo{person}{Jian{-}Yun Nie}, {and} \bibinfo{person}{Ji{-}Rong Wen}.} \bibinfo{year}{2023}\natexlab{a}.
\newblock \showarticletitle{HaluEval: {A} Large-Scale Hallucination Evaluation Benchmark for Large Language Models}. In \bibinfo{booktitle}{\emph{Proceedings of the 2023 Conference on Empirical Methods in Natural Language Processing, {EMNLP} 2023, Singapore, December 6-10, 2023}}, \bibfield{editor}{\bibinfo{person}{Houda Bouamor}, \bibinfo{person}{Juan Pino}, {and} \bibinfo{person}{Kalika Bali}} (Eds.). \bibinfo{publisher}{Association for Computational Linguistics}, \bibinfo{pages}{6449--6464}.
\newblock
\urldef\tempurl%
\url{https://doi.org/10.18653/V1/2023.EMNLP-MAIN.397}
\showDOI{\tempurl}


\bibitem[Li et~al\mbox{.}(2023b)]%
        {li2023inference}
\bibfield{author}{\bibinfo{person}{Kenneth Li}, \bibinfo{person}{Oam Patel}, \bibinfo{person}{Fernanda~B. Vi{\'{e}}gas}, \bibinfo{person}{Hanspeter Pfister}, {and} \bibinfo{person}{Martin Wattenberg}.} \bibinfo{year}{2023}\natexlab{b}.
\newblock \showarticletitle{Inference-Time Intervention: Eliciting Truthful Answers from a Language Model}. In \bibinfo{booktitle}{\emph{Advances in Neural Information Processing Systems 36: Annual Conference on Neural Information Processing Systems 2023, NeurIPS 2023, New Orleans, LA, USA, December 10 - 16, 2023}}, \bibfield{editor}{\bibinfo{person}{Alice Oh}, \bibinfo{person}{Tristan Naumann}, \bibinfo{person}{Amir Globerson}, \bibinfo{person}{Kate Saenko}, \bibinfo{person}{Moritz Hardt}, {and} \bibinfo{person}{Sergey Levine}} (Eds.).
\newblock
\urldef\tempurl%
\url{http://papers.nips.cc/paper\_files/paper/2023/hash/81b8390039b7302c909cb769f8b6cd93-Abstract-Conference.html}
\showURL{%
\tempurl}


\bibitem[Liang et~al\mbox{.}(2022)]%
        {liang2022reasoning}
\bibfield{author}{\bibinfo{person}{Ke Liang}, \bibinfo{person}{Lingyuan Meng}, \bibinfo{person}{Meng Liu}, \bibinfo{person}{Yue Liu}, \bibinfo{person}{Wenxuan Tu}, \bibinfo{person}{Siwei Wang}, \bibinfo{person}{Sihang Zhou}, \bibinfo{person}{Xinwang Liu}, {and} \bibinfo{person}{Fuchun Sun}.} \bibinfo{year}{2022}\natexlab{}.
\newblock \showarticletitle{Reasoning over different types of knowledge graphs: Static, temporal and multi-modal}.
\newblock \bibinfo{journal}{\emph{arXiv preprint arXiv:2212.05767}} (\bibinfo{year}{2022}).
\newblock


\bibitem[Lightman et~al\mbox{.}(2024)]%
        {lightman2023let}
\bibfield{author}{\bibinfo{person}{Hunter Lightman}, \bibinfo{person}{Vineet Kosaraju}, \bibinfo{person}{Yuri Burda}, \bibinfo{person}{Harrison Edwards}, \bibinfo{person}{Bowen Baker}, \bibinfo{person}{Teddy Lee}, \bibinfo{person}{Jan Leike}, \bibinfo{person}{John Schulman}, \bibinfo{person}{Ilya Sutskever}, {and} \bibinfo{person}{Karl Cobbe}.} \bibinfo{year}{2024}\natexlab{}.
\newblock \showarticletitle{Let's Verify Step by Step}. In \bibinfo{booktitle}{\emph{The Twelfth International Conference on Learning Representations, {ICLR} 2024, Vienna, Austria, May 7-11, 2024}}. \bibinfo{publisher}{OpenReview.net}.
\newblock
\urldef\tempurl%
\url{https://openreview.net/forum?id=v8L0pN6EOi}
\showURL{%
\tempurl}


\bibitem[Lin et~al\mbox{.}(2022)]%
        {lin-etal-2022-truthfulqa}
\bibfield{author}{\bibinfo{person}{Stephanie Lin}, \bibinfo{person}{Jacob Hilton}, {and} \bibinfo{person}{Owain Evans}.} \bibinfo{year}{2022}\natexlab{}.
\newblock \showarticletitle{{T}ruthful{QA}: Measuring How Models Mimic Human Falsehoods}. In \bibinfo{booktitle}{\emph{Proceedings of the 60th Annual Meeting of the Association for Computational Linguistics (Volume 1: Long Papers)}}, \bibfield{editor}{\bibinfo{person}{Smaranda Muresan}, \bibinfo{person}{Preslav Nakov}, {and} \bibinfo{person}{Aline Villavicencio}} (Eds.). \bibinfo{publisher}{Association for Computational Linguistics}, \bibinfo{address}{Dublin, Ireland}, \bibinfo{pages}{3214--3252}.
\newblock
\urldef\tempurl%
\url{https://doi.org/10.18653/v1/2022.acl-long.229}
\showDOI{\tempurl}


\bibitem[Meng et~al\mbox{.}(2022)]%
        {meng2022locating}
\bibfield{author}{\bibinfo{person}{Kevin Meng}, \bibinfo{person}{David Bau}, \bibinfo{person}{Alex Andonian}, {and} \bibinfo{person}{Yonatan Belinkov}.} \bibinfo{year}{2022}\natexlab{}.
\newblock \showarticletitle{Locating and Editing Factual Associations in {GPT}}.
\newblock \bibinfo{journal}{\emph{Advances in Neural Information Processing Systems}}  \bibinfo{volume}{35} (\bibinfo{year}{2022}).
\newblock


\bibitem[Meng et~al\mbox{.}(2023)]%
        {meng2022memit}
\bibfield{author}{\bibinfo{person}{Kevin Meng}, \bibinfo{person}{Arnab~Sen Sharma}, \bibinfo{person}{Alex~J. Andonian}, \bibinfo{person}{Yonatan Belinkov}, {and} \bibinfo{person}{David Bau}.} \bibinfo{year}{2023}\natexlab{}.
\newblock \showarticletitle{Mass-Editing Memory in a Transformer}. In \bibinfo{booktitle}{\emph{The Eleventh International Conference on Learning Representations, {ICLR} 2023, Kigali, Rwanda, May 1-5, 2023}}. \bibinfo{publisher}{OpenReview.net}.
\newblock
\urldef\tempurl%
\url{https://openreview.net/forum?id=MkbcAHIYgyS}
\showURL{%
\tempurl}


\bibitem[Miller(1994)]%
        {miller-1994-wordnet}
\bibfield{author}{\bibinfo{person}{George~A. Miller}.} \bibinfo{year}{1994}\natexlab{}.
\newblock \showarticletitle{{W}ord{N}et: A Lexical Database for {E}nglish}. In \bibinfo{booktitle}{\emph{{H}uman {L}anguage {T}echnology: Proceedings of a Workshop held at {P}lainsboro, {N}ew {J}ersey, {M}arch 8-11, 1994}}.
\newblock
\urldef\tempurl%
\url{https://aclanthology.org/H94-1111}
\showURL{%
\tempurl}


\bibitem[Miller(1995)]%
        {WordNet}
\bibfield{author}{\bibinfo{person}{George~A. Miller}.} \bibinfo{year}{1995}\natexlab{}.
\newblock \showarticletitle{WordNet: A Lexical Database for English}.
\newblock \bibinfo{journal}{\emph{Commun. ACM}} \bibinfo{volume}{38}, \bibinfo{number}{11} (\bibinfo{date}{nov} \bibinfo{year}{1995}), \bibinfo{pages}{39–41}.
\newblock
\showISSN{0001-0782}
\urldef\tempurl%
\url{https://doi.org/10.1145/219717.219748}
\showDOI{\tempurl}


\bibitem[Min et~al\mbox{.}(2023)]%
        {min2023factscore}
\bibfield{author}{\bibinfo{person}{Sewon Min}, \bibinfo{person}{Kalpesh Krishna}, \bibinfo{person}{Xinxi Lyu}, \bibinfo{person}{Mike Lewis}, \bibinfo{person}{Wen{-}tau Yih}, \bibinfo{person}{Pang~Wei Koh}, \bibinfo{person}{Mohit Iyyer}, \bibinfo{person}{Luke Zettlemoyer}, {and} \bibinfo{person}{Hannaneh Hajishirzi}.} \bibinfo{year}{2023}\natexlab{}.
\newblock \showarticletitle{FActScore: Fine-grained Atomic Evaluation of Factual Precision in Long Form Text Generation}. In \bibinfo{booktitle}{\emph{Proceedings of the 2023 Conference on Empirical Methods in Natural Language Processing, {EMNLP} 2023, Singapore, December 6-10, 2023}}, \bibfield{editor}{\bibinfo{person}{Houda Bouamor}, \bibinfo{person}{Juan Pino}, {and} \bibinfo{person}{Kalika Bali}} (Eds.). \bibinfo{publisher}{Association for Computational Linguistics}, \bibinfo{pages}{12076--12100}.
\newblock
\urldef\tempurl%
\url{https://doi.org/10.18653/V1/2023.EMNLP-MAIN.741}
\showDOI{\tempurl}


\bibitem[Moon et~al\mbox{.}(2019)]%
        {moon2019opendialkg}
\bibfield{author}{\bibinfo{person}{Seungwhan Moon}, \bibinfo{person}{Pararth Shah}, \bibinfo{person}{Anuj Kumar}, {and} \bibinfo{person}{Rajen Subba}.} \bibinfo{year}{2019}\natexlab{}.
\newblock \showarticletitle{Opendialkg: Explainable conversational reasoning with attention-based walks over knowledge graphs}. In \bibinfo{booktitle}{\emph{Proceedings of the 57th annual meeting of the association for computational linguistics}}. \bibinfo{pages}{845--854}.
\newblock


\bibitem[Olausson et~al\mbox{.}(2023)]%
        {olausson-etal-2023-linc}
\bibfield{author}{\bibinfo{person}{Theo Olausson}, \bibinfo{person}{Alex Gu}, \bibinfo{person}{Ben Lipkin}, \bibinfo{person}{Cedegao Zhang}, \bibinfo{person}{Armando Solar-Lezama}, \bibinfo{person}{Joshua Tenenbaum}, {and} \bibinfo{person}{Roger Levy}.} \bibinfo{year}{2023}\natexlab{}.
\newblock \showarticletitle{{LINC}: A Neurosymbolic Approach for Logical Reasoning by Combining Language Models with First-Order Logic Provers}. In \bibinfo{booktitle}{\emph{Proceedings of the 2023 Conference on Empirical Methods in Natural Language Processing}}, \bibfield{editor}{\bibinfo{person}{Houda Bouamor}, \bibinfo{person}{Juan Pino}, {and} \bibinfo{person}{Kalika Bali}} (Eds.). \bibinfo{publisher}{Association for Computational Linguistics}, \bibinfo{address}{Singapore}, \bibinfo{pages}{5153--5176}.
\newblock
\urldef\tempurl%
\url{https://doi.org/10.18653/v1/2023.emnlp-main.313}
\showDOI{\tempurl}


\bibitem[OpenAI(2023)]%
        {OpenAI2023GPT4TR}
\bibfield{author}{\bibinfo{person}{OpenAI}.} \bibinfo{year}{2023}\natexlab{}.
\newblock \showarticletitle{GPT-4 Technical Report}.
\newblock \bibinfo{journal}{\emph{ArXiv}}  \bibinfo{volume}{abs/2303.08774} (\bibinfo{year}{2023}).
\newblock


\bibitem[Pal et~al\mbox{.}(2023)]%
        {umapathi2023med}
\bibfield{author}{\bibinfo{person}{Ankit Pal}, \bibinfo{person}{Logesh~Kumar Umapathi}, {and} \bibinfo{person}{Malaikannan Sankarasubbu}.} \bibinfo{year}{2023}\natexlab{}.
\newblock \showarticletitle{Med-HALT: Medical Domain Hallucination Test for Large Language Models}. In \bibinfo{booktitle}{\emph{Proceedings of the 27th Conference on Computational Natural Language Learning, CoNLL 2023, Singapore, December 6-7, 2023}}, \bibfield{editor}{\bibinfo{person}{Jing Jiang}, \bibinfo{person}{David Reitter}, {and} \bibinfo{person}{Shumin Deng}} (Eds.). \bibinfo{publisher}{Association for Computational Linguistics}, \bibinfo{pages}{314--334}.
\newblock
\urldef\tempurl%
\url{https://doi.org/10.18653/V1/2023.CONLL-1.21}
\showDOI{\tempurl}


\bibitem[Pan et~al\mbox{.}(2023)]%
        {pan-etal-2023-logic}
\bibfield{author}{\bibinfo{person}{Liangming Pan}, \bibinfo{person}{Alon Albalak}, \bibinfo{person}{Xinyi Wang}, {and} \bibinfo{person}{William Wang}.} \bibinfo{year}{2023}\natexlab{}.
\newblock \showarticletitle{Logic-{LM}: Empowering Large Language Models with Symbolic Solvers for Faithful Logical Reasoning}. In \bibinfo{booktitle}{\emph{Findings of the Association for Computational Linguistics: EMNLP 2023}}, \bibfield{editor}{\bibinfo{person}{Houda Bouamor}, \bibinfo{person}{Juan Pino}, {and} \bibinfo{person}{Kalika Bali}} (Eds.). \bibinfo{publisher}{Association for Computational Linguistics}, \bibinfo{address}{Singapore}, \bibinfo{pages}{3806--3824}.
\newblock
\urldef\tempurl%
\url{https://doi.org/10.18653/v1/2023.findings-emnlp.248}
\showDOI{\tempurl}


\bibitem[Prud'hommeaux and Seaborne(2018)]%
        {sparql}
\bibfield{author}{\bibinfo{person}{Eric Prud'hommeaux} {and} \bibinfo{person}{Andy Seaborne}.} \bibinfo{year}{2018}\natexlab{}.
\newblock \bibinfo{title}{{SPARQL Query Language for RDF - W3C recommendation.}}
\newblock \bibinfo{howpublished}{\url{https://www.w3.org/TR/rdf-sparql-query/}}.
\newblock


\bibitem[Qiu et~al\mbox{.}(2024)]%
        {qiu2023large}
\bibfield{author}{\bibinfo{person}{Yifu Qiu}, \bibinfo{person}{Zheng Zhao}, \bibinfo{person}{Yftah Ziser}, \bibinfo{person}{Anna Korhonen}, \bibinfo{person}{Edoardo Ponti}, {and} \bibinfo{person}{Shay Cohen}.} \bibinfo{year}{2024}\natexlab{}.
\newblock \showarticletitle{Are Large Language Model Temporally Grounded?}. In \bibinfo{booktitle}{\emph{Proceedings of the 2024 Conference of the North American Chapter of the Association for Computational Linguistics: Human Language Technologies (Volume 1: Long Papers)}}, \bibfield{editor}{\bibinfo{person}{Kevin Duh}, \bibinfo{person}{Helena Gomez}, {and} \bibinfo{person}{Steven Bethard}} (Eds.). \bibinfo{publisher}{Association for Computational Linguistics}, \bibinfo{address}{Mexico City, Mexico}, \bibinfo{pages}{7064--7083}.
\newblock
\urldef\tempurl%
\url{https://doi.org/10.18653/v1/2024.naacl-long.391}
\showDOI{\tempurl}


\bibitem[Reimers and Gurevych(2019)]%
        {reimers2019sentence}
\bibfield{author}{\bibinfo{person}{Nils Reimers} {and} \bibinfo{person}{Iryna Gurevych}.} \bibinfo{year}{2019}\natexlab{}.
\newblock \showarticletitle{Sentence-BERT: Sentence Embeddings using Siamese BERT-Networks}. In \bibinfo{booktitle}{\emph{Proceedings of the 2019 Conference on Empirical Methods in Natural Language Processing and the 9th International Joint Conference on Natural Language Processing, {EMNLP-IJCNLP} 2019, Hong Kong, China, November 3-7, 2019}}, \bibfield{editor}{\bibinfo{person}{Kentaro Inui}, \bibinfo{person}{Jing Jiang}, \bibinfo{person}{Vincent Ng}, {and} \bibinfo{person}{Xiaojun Wan}} (Eds.). \bibinfo{publisher}{Association for Computational Linguistics}, \bibinfo{pages}{3980--3990}.
\newblock
\urldef\tempurl%
\url{https://doi.org/10.18653/V1/D19-1410}
\showDOI{\tempurl}


\bibitem[Remy(2020)]%
        {StanfordOpenIEWrapper}
\bibfield{author}{\bibinfo{person}{Philippe Remy}.} \bibinfo{year}{2020}\natexlab{}.
\newblock \bibinfo{title}{Python wrapper for Stanford OpenIE}.
\newblock \bibinfo{howpublished}{\url{https://github.com/philipperemy/Stanford-OpenIE-Python}}.
\newblock


\bibitem[Ren and Leskovec(2020)]%
        {ren2020beta}
\bibfield{author}{\bibinfo{person}{Hongyu Ren} {and} \bibinfo{person}{Jure Leskovec}.} \bibinfo{year}{2020}\natexlab{}.
\newblock \showarticletitle{Beta embeddings for multi-hop logical reasoning in knowledge graphs}.
\newblock \bibinfo{journal}{\emph{Advances in Neural Information Processing Systems}}  \bibinfo{volume}{33} (\bibinfo{year}{2020}), \bibinfo{pages}{19716--19726}.
\newblock


\bibitem[{Satoshi Tajiri}(2023)]%
        {pokemon}
\bibfield{author}{\bibinfo{person}{{Satoshi Tajiri}}.} \bibinfo{year}{2023}\natexlab{}.
\newblock \bibinfo{title}{{Pokemon}}.
\newblock \bibinfo{howpublished}{\url{https://www.pokemon.com/us}}.
\newblock


\bibitem[{ScienceDirect}(2023)]%
        {J_S}
\bibfield{author}{\bibinfo{person}{{ScienceDirect}}.} \bibinfo{year}{2023}\natexlab{}.
\newblock \bibinfo{title}{{Jaccard Similarity}}.
\newblock \bibinfo{howpublished}{\url{https://www.sciencedirect.com/topics/computer-science/jaccard-similarity}}.
\newblock


\bibitem[{ScienceDirect}(2024)]%
        {F_T}
\bibfield{author}{\bibinfo{person}{{ScienceDirect}}.} \bibinfo{year}{2024}\natexlab{}.
\newblock \bibinfo{title}{{Friedman Test}}.
\newblock \bibinfo{howpublished}{\url{https://www.sciencedirect.com/topics/biochemistry-genetics-and-molecular-biology/friedman-test}}.
\newblock


\bibitem[See et~al\mbox{.}(2017)]%
        {see2017get}
\bibfield{author}{\bibinfo{person}{Abigail See}, \bibinfo{person}{Peter~J. Liu}, {and} \bibinfo{person}{Christopher~D. Manning}.} \bibinfo{year}{2017}\natexlab{}.
\newblock \showarticletitle{Get To The Point: Summarization with Pointer-Generator Networks}. In \bibinfo{booktitle}{\emph{Proceedings of the 55th Annual Meeting of the Association for Computational Linguistics (Volume 1: Long Papers)}}, \bibfield{editor}{\bibinfo{person}{Regina Barzilay} {and} \bibinfo{person}{Min-Yen Kan}} (Eds.). \bibinfo{publisher}{Association for Computational Linguistics}, \bibinfo{address}{Vancouver, Canada}, \bibinfo{pages}{1073--1083}.
\newblock
\urldef\tempurl%
\url{https://doi.org/10.18653/v1/P17-1099}
\showDOI{\tempurl}


\bibitem[Siddiq and Santos(2023)]%
        {siddiq2023generate}
\bibfield{author}{\bibinfo{person}{Mohammed~Latif Siddiq} {and} \bibinfo{person}{Joanna Santos}.} \bibinfo{year}{2023}\natexlab{}.
\newblock \showarticletitle{Generate and pray: Using sallms to evaluate the security of llm generated code}.
\newblock \bibinfo{journal}{\emph{arXiv preprint arXiv:2311.00889}} (\bibinfo{year}{2023}).
\newblock


\bibitem[Suchanek et~al\mbox{.}(2007)]%
        {Yago}
\bibfield{author}{\bibinfo{person}{Fabian~M. Suchanek}, \bibinfo{person}{Gjergji Kasneci}, {and} \bibinfo{person}{Gerhard Weikum}.} \bibinfo{year}{2007}\natexlab{}.
\newblock \showarticletitle{Yago: A Core of Semantic Knowledge}. In \bibinfo{booktitle}{\emph{Proceedings of the 16th International Conference on World Wide Web}} (Banff, Alberta, Canada) \emph{(\bibinfo{series}{WWW '07})}. \bibinfo{publisher}{Association for Computing Machinery}, \bibinfo{address}{New York, NY, USA}, \bibinfo{pages}{697–706}.
\newblock
\showISBNx{9781595936547}
\urldef\tempurl%
\url{https://doi.org/10.1145/1242572.1242667}
\showDOI{\tempurl}


\bibitem[Tian et~al\mbox{.}(2024)]%
        {tian2023fine}
\bibfield{author}{\bibinfo{person}{Katherine Tian}, \bibinfo{person}{Eric Mitchell}, \bibinfo{person}{Huaxiu Yao}, \bibinfo{person}{Christopher~D. Manning}, {and} \bibinfo{person}{Chelsea Finn}.} \bibinfo{year}{2024}\natexlab{}.
\newblock \showarticletitle{Fine-Tuning Language Models for Factuality}. In \bibinfo{booktitle}{\emph{The Twelfth International Conference on Learning Representations, {ICLR} 2024, Vienna, Austria, May 7-11, 2024}}. \bibinfo{publisher}{OpenReview.net}.
\newblock
\urldef\tempurl%
\url{https://openreview.net/forum?id=WPZ2yPag4K}
\showURL{%
\tempurl}


\bibitem[Tian et~al\mbox{.}(2022)]%
        {TIAN2022100159}
\bibfield{author}{\bibinfo{person}{Ling Tian}, \bibinfo{person}{Xue Zhou}, \bibinfo{person}{Yan-Ping Wu}, \bibinfo{person}{Wang-Tao Zhou}, \bibinfo{person}{Jin-Hao Zhang}, {and} \bibinfo{person}{Tian-Shu Zhang}.} \bibinfo{year}{2022}\natexlab{}.
\newblock \showarticletitle{Knowledge graph and knowledge reasoning: A systematic review}.
\newblock \bibinfo{journal}{\emph{Journal of Electronic Science and Technology}} \bibinfo{volume}{20}, \bibinfo{number}{2} (\bibinfo{year}{2022}), \bibinfo{pages}{100159}.
\newblock
\showISSN{1674-862X}
\urldef\tempurl%
\url{https://doi.org/10.1016/j.jnlest.2022.100159}
\showDOI{\tempurl}


\bibitem[Touvron et~al\mbox{.}(2023)]%
        {touvron2023llama}
\bibfield{author}{\bibinfo{person}{Hugo Touvron}, \bibinfo{person}{Louis Martin}, \bibinfo{person}{Kevin Stone}, \bibinfo{person}{Peter Albert}, \bibinfo{person}{Amjad Almahairi}, \bibinfo{person}{Yasmine Babaei}, \bibinfo{person}{Nikolay Bashlykov}, \bibinfo{person}{Soumya Batra}, \bibinfo{person}{Prajjwal Bhargava}, \bibinfo{person}{Shruti Bhosale}, {et~al\mbox{.}}} \bibinfo{year}{2023}\natexlab{}.
\newblock \showarticletitle{Llama 2: Open foundation and fine-tuned chat models}.
\newblock \bibinfo{journal}{\emph{arXiv preprint arXiv:2307.09288}} (\bibinfo{year}{2023}).
\newblock


\bibitem[Varshney et~al\mbox{.}(2023)]%
        {varshney2023stitch}
\bibfield{author}{\bibinfo{person}{Neeraj Varshney}, \bibinfo{person}{Wenlin Yao}, \bibinfo{person}{Hongming Zhang}, \bibinfo{person}{Jianshu Chen}, {and} \bibinfo{person}{Dong Yu}.} \bibinfo{year}{2023}\natexlab{}.
\newblock \showarticletitle{A stitch in time saves nine: Detecting and mitigating hallucinations of llms by validating low-confidence generation}.
\newblock \bibinfo{journal}{\emph{arXiv preprint arXiv:2307.03987}} (\bibinfo{year}{2023}).
\newblock


\bibitem[Vu et~al\mbox{.}(2024)]%
        {vu2023freshllms}
\bibfield{author}{\bibinfo{person}{Tu Vu}, \bibinfo{person}{Mohit Iyyer}, \bibinfo{person}{Xuezhi Wang}, \bibinfo{person}{Noah Constant}, \bibinfo{person}{Jerry~W. Wei}, \bibinfo{person}{Jason Wei}, \bibinfo{person}{Chris Tar}, \bibinfo{person}{Yun{-}Hsuan Sung}, \bibinfo{person}{Denny Zhou}, \bibinfo{person}{Quoc~V. Le}, {and} \bibinfo{person}{Thang Luong}.} \bibinfo{year}{2024}\natexlab{}.
\newblock \showarticletitle{FreshLLMs: Refreshing Large Language Models with Search Engine Augmentation}. In \bibinfo{booktitle}{\emph{Findings of the Association for Computational Linguistics, {ACL} 2024, Bangkok, Thailand and virtual meeting, August 11-16, 2024}}, \bibfield{editor}{\bibinfo{person}{Lun{-}Wei Ku}, \bibinfo{person}{Andre Martins}, {and} \bibinfo{person}{Vivek Srikumar}} (Eds.). \bibinfo{publisher}{Association for Computational Linguistics}, \bibinfo{pages}{13697--13720}.
\newblock
\urldef\tempurl%
\url{https://aclanthology.org/2024.findings-acl.813}
\showURL{%
\tempurl}


\bibitem[Wang et~al\mbox{.}(2024)]%
        {wang2023easyedit}
\bibfield{author}{\bibinfo{person}{Peng Wang}, \bibinfo{person}{Ningyu Zhang}, \bibinfo{person}{Bozhong Tian}, \bibinfo{person}{Zekun Xi}, \bibinfo{person}{Yunzhi Yao}, \bibinfo{person}{Ziwen Xu}, \bibinfo{person}{Mengru Wang}, \bibinfo{person}{Shengyu Mao}, \bibinfo{person}{Xiaohan Wang}, \bibinfo{person}{Siyuan Cheng}, \bibinfo{person}{Kangwei Liu}, \bibinfo{person}{Yuansheng Ni}, \bibinfo{person}{Guozhou Zheng}, {and} \bibinfo{person}{Huajun Chen}.} \bibinfo{year}{2024}\natexlab{}.
\newblock \showarticletitle{{E}asy{E}dit: An Easy-to-use Knowledge Editing Framework for Large Language Models}. In \bibinfo{booktitle}{\emph{Proceedings of the 62nd Annual Meeting of the Association for Computational Linguistics (Volume 3: System Demonstrations)}}, \bibfield{editor}{\bibinfo{person}{Yixin Cao}, \bibinfo{person}{Yang Feng}, {and} \bibinfo{person}{Deyi Xiong}} (Eds.). \bibinfo{publisher}{Association for Computational Linguistics}, \bibinfo{address}{Bangkok, Thailand}, \bibinfo{pages}{82--93}.
\newblock
\urldef\tempurl%
\url{https://aclanthology.org/2024.acl-demos.9}
\showURL{%
\tempurl}


\bibitem[Wang et~al\mbox{.}(2021)]%
        {phrasebertwang2021}
\bibfield{author}{\bibinfo{person}{Shufan Wang}, \bibinfo{person}{Laure Thompson}, {and} \bibinfo{person}{Mohit Iyyer}.} \bibinfo{year}{2021}\natexlab{}.
\newblock \showarticletitle{Phrase-BERT: Improved Phrase Embeddings from BERT with an Application to Corpus Exploration}. In \bibinfo{booktitle}{\emph{Empirical Methods in Natural Language Processing}}.
\newblock


\bibitem[Wielemaker et~al\mbox{.}(2012)]%
        {wielemaker2012swi}
\bibfield{author}{\bibinfo{person}{Jan Wielemaker}, \bibinfo{person}{Tom Schrijvers}, \bibinfo{person}{Markus Triska}, {and} \bibinfo{person}{Torbj{\"o}rn Lager}.} \bibinfo{year}{2012}\natexlab{}.
\newblock \showarticletitle{Swi-prolog}.
\newblock \bibinfo{journal}{\emph{Theory and Practice of Logic Programming}} \bibinfo{volume}{12}, \bibinfo{number}{1-2} (\bibinfo{year}{2012}), \bibinfo{pages}{67--96}.
\newblock


\bibitem[Xu et~al\mbox{.}(2024)]%
        {xu2024largelanguagemodelscyber}
\bibfield{author}{\bibinfo{person}{Hanxiang Xu}, \bibinfo{person}{Shenao Wang}, \bibinfo{person}{Ningke Li}, \bibinfo{person}{Kailong Wang}, \bibinfo{person}{Yanjie Zhao}, \bibinfo{person}{Kai Chen}, \bibinfo{person}{Ting Yu}, \bibinfo{person}{Yang Liu}, {and} \bibinfo{person}{Haoyu Wang}.} \bibinfo{year}{2024}\natexlab{}.
\newblock \bibinfo{title}{Large Language Models for Cyber Security: A Systematic Literature Review}.
\newblock
\newblock
\showeprint[arxiv]{2405.04760}~[cs.CR]
\urldef\tempurl%
\url{https://arxiv.org/abs/2405.04760}
\showURL{%
\tempurl}


\bibitem[Yang et~al\mbox{.}(2024a)]%
        {distillseq}
\bibfield{author}{\bibinfo{person}{Mingke Yang}, \bibinfo{person}{Yuqi Chen}, \bibinfo{person}{Yi Liu}, {and} \bibinfo{person}{Ling Shi}.} \bibinfo{year}{2024}\natexlab{a}.
\newblock \showarticletitle{DistillSeq: A Framework for Safety Alignment Testing in Large Language Models using Knowledge Distillatio}. In \bibinfo{booktitle}{\emph{The ACM SIGSOFT International Symposium on Software Testing and Analysis}}.
\newblock


\bibitem[Yang et~al\mbox{.}(2024b)]%
        {yang-etal-2024-language}
\bibfield{author}{\bibinfo{person}{Zonglin Yang}, \bibinfo{person}{Li Dong}, \bibinfo{person}{Xinya Du}, \bibinfo{person}{Hao Cheng}, \bibinfo{person}{Erik Cambria}, \bibinfo{person}{Xiaodong Liu}, \bibinfo{person}{Jianfeng Gao}, {and} \bibinfo{person}{Furu Wei}.} \bibinfo{year}{2024}\natexlab{b}.
\newblock \showarticletitle{Language Models as Inductive Reasoners}. In \bibinfo{booktitle}{\emph{Proceedings of the 18th Conference of the European Chapter of the Association for Computational Linguistics (Volume 1: Long Papers)}}, \bibfield{editor}{\bibinfo{person}{Yvette Graham} {and} \bibinfo{person}{Matthew Purver}} (Eds.). \bibinfo{publisher}{Association for Computational Linguistics}, \bibinfo{address}{St. Julian{'}s, Malta}, \bibinfo{pages}{209--225}.
\newblock
\urldef\tempurl%
\url{https://aclanthology.org/2024.eacl-long.13}
\showURL{%
\tempurl}


\bibitem[Yang et~al\mbox{.}(2018)]%
        {yang2018hotpotqa}
\bibfield{author}{\bibinfo{person}{Zhilin Yang}, \bibinfo{person}{Peng Qi}, \bibinfo{person}{Saizheng Zhang}, \bibinfo{person}{Yoshua Bengio}, \bibinfo{person}{William~W. Cohen}, \bibinfo{person}{Ruslan Salakhutdinov}, {and} \bibinfo{person}{Christopher~D. Manning}.} \bibinfo{year}{2018}\natexlab{}.
\newblock \showarticletitle{HotpotQA: {A} Dataset for Diverse, Explainable Multi-hop Question Answering}. In \bibinfo{booktitle}{\emph{Proceedings of the 2018 Conference on Empirical Methods in Natural Language Processing, Brussels, Belgium, October 31 - November 4, 2018}}, \bibfield{editor}{\bibinfo{person}{Ellen Riloff}, \bibinfo{person}{David Chiang}, \bibinfo{person}{Julia Hockenmaier}, {and} \bibinfo{person}{Jun'ichi Tsujii}} (Eds.). \bibinfo{publisher}{Association for Computational Linguistics}, \bibinfo{pages}{2369--2380}.
\newblock
\urldef\tempurl%
\url{https://doi.org/10.18653/V1/D18-1259}
\showDOI{\tempurl}


\bibitem[Yao et~al\mbox{.}(2024)]%
        {yao2023survey}
\bibfield{author}{\bibinfo{person}{Yifan Yao}, \bibinfo{person}{Jinhao Duan}, \bibinfo{person}{Kaidi Xu}, \bibinfo{person}{Yuanfang Cai}, \bibinfo{person}{Zhibo Sun}, {and} \bibinfo{person}{Yue Zhang}.} \bibinfo{year}{2024}\natexlab{}.
\newblock \showarticletitle{A survey on large language model (LLM) security and privacy: The Good, The Bad, and The Ugly}.
\newblock \bibinfo{journal}{\emph{High-Confidence Computing}} \bibinfo{volume}{4}, \bibinfo{number}{2} (\bibinfo{year}{2024}), \bibinfo{pages}{100211}.
\newblock
\showISSN{2667-2952}
\urldef\tempurl%
\url{https://doi.org/10.1016/j.hcc.2024.100211}
\showDOI{\tempurl}


\bibitem[Ye et~al\mbox{.}(2023)]%
        {Ye-Et-Al:2023:SAT}
\bibfield{author}{\bibinfo{person}{Xi Ye}, \bibinfo{person}{Qiaochu Chen}, \bibinfo{person}{Isil Dillig}, {and} \bibinfo{person}{Greg Durrett}.} \bibinfo{year}{2023}\natexlab{}.
\newblock \showarticletitle{SatLM: Satisfiability-Aided Language Models Using Declarative Prompting}. In \bibinfo{booktitle}{\emph{Advances in Neural Information Processing Systems 36: Annual Conference on Neural Information Processing Systems 2023, NeurIPS 2023, New Orleans, LA, USA, December 10 - 16, 2023}}, \bibfield{editor}{\bibinfo{person}{Alice Oh}, \bibinfo{person}{Tristan Naumann}, \bibinfo{person}{Amir Globerson}, \bibinfo{person}{Kate Saenko}, \bibinfo{person}{Moritz Hardt}, {and} \bibinfo{person}{Sergey Levine}} (Eds.).
\newblock
\urldef\tempurl%
\url{http://papers.nips.cc/paper\_files/paper/2023/hash/8e9c7d4a48bdac81a58f983a64aaf42b-Abstract-Conference.html}
\showURL{%
\tempurl}


\bibitem[Yin et~al\mbox{.}(2023)]%
        {yin2023woodpecker}
\bibfield{author}{\bibinfo{person}{Shukang Yin}, \bibinfo{person}{Chaoyou Fu}, \bibinfo{person}{Sirui Zhao}, \bibinfo{person}{Tong Xu}, \bibinfo{person}{Hao Wang}, \bibinfo{person}{Dianbo Sui}, \bibinfo{person}{Yunhang Shen}, \bibinfo{person}{Ke Li}, \bibinfo{person}{Xing Sun}, {and} \bibinfo{person}{Enhong Chen}.} \bibinfo{year}{2023}\natexlab{}.
\newblock \showarticletitle{Woodpecker: Hallucination correction for multimodal large language models}.
\newblock \bibinfo{journal}{\emph{arXiv preprint arXiv:2310.16045}} (\bibinfo{year}{2023}).
\newblock


\bibitem[Yu et~al\mbox{.}(2024)]%
        {yu2023kola}
\bibfield{author}{\bibinfo{person}{Jifan Yu}, \bibinfo{person}{Xiaozhi Wang}, \bibinfo{person}{Shangqing Tu}, \bibinfo{person}{Shulin Cao}, \bibinfo{person}{Daniel Zhang{-}Li}, \bibinfo{person}{Xin Lv}, \bibinfo{person}{Hao Peng}, \bibinfo{person}{Zijun Yao}, \bibinfo{person}{Xiaohan Zhang}, \bibinfo{person}{Hanming Li}, \bibinfo{person}{Chunyang Li}, \bibinfo{person}{Zheyuan Zhang}, \bibinfo{person}{Yushi Bai}, \bibinfo{person}{Yantao Liu}, \bibinfo{person}{Amy Xin}, \bibinfo{person}{Kaifeng Yun}, \bibinfo{person}{Linlu Gong}, \bibinfo{person}{Nianyi Lin}, \bibinfo{person}{Jianhui Chen}, \bibinfo{person}{Zhili Wu}, \bibinfo{person}{Yunjia Qi}, \bibinfo{person}{Weikai Li}, \bibinfo{person}{Yong Guan}, \bibinfo{person}{Kaisheng Zeng}, \bibinfo{person}{Ji Qi}, \bibinfo{person}{Hailong Jin}, \bibinfo{person}{Jinxin Liu}, \bibinfo{person}{Yu Gu}, \bibinfo{person}{Yuan Yao}, \bibinfo{person}{Ning Ding}, \bibinfo{person}{Lei Hou}, \bibinfo{person}{Zhiyuan Liu}, \bibinfo{person}{Bin Xu}, \bibinfo{person}{Jie Tang},
  {and} \bibinfo{person}{Juanzi Li}.} \bibinfo{year}{2024}\natexlab{}.
\newblock \showarticletitle{KoLA: Carefully Benchmarking World Knowledge of Large Language Models}. In \bibinfo{booktitle}{\emph{The Twelfth International Conference on Learning Representations, {ICLR} 2024, Vienna, Austria, May 7-11, 2024}}. \bibinfo{publisher}{OpenReview.net}.
\newblock
\urldef\tempurl%
\url{https://openreview.net/forum?id=AqN23oqraW}
\showURL{%
\tempurl}


\bibitem[Zhang et~al\mbox{.}(2023)]%
        {zhang2023hallucination}
\bibfield{author}{\bibinfo{person}{Yue Zhang}, \bibinfo{person}{Yafu Li}, \bibinfo{person}{Leyang Cui}, \bibinfo{person}{Deng Cai}, \bibinfo{person}{Lemao Liu}, \bibinfo{person}{Tingchen Fu}, \bibinfo{person}{Xinting Huang}, \bibinfo{person}{Enbo Zhao}, \bibinfo{person}{Yu Zhang}, \bibinfo{person}{Yulong Chen}, \bibinfo{person}{Longyue Wang}, \bibinfo{person}{Anh~Tuan Luu}, \bibinfo{person}{Wei Bi}, \bibinfo{person}{Freda Shi}, {and} \bibinfo{person}{Shuming Shi}.} \bibinfo{year}{2023}\natexlab{}.
\newblock \showarticletitle{Siren's Song in the AI Ocean: A Survey on Hallucination in Large Language Models}.
\newblock \bibinfo{journal}{\emph{arXiv preprint arXiv:2309.01219}} (\bibinfo{year}{2023}).
\newblock


\bibitem[Zhang et~al\mbox{.}(2024)]%
        {zhang2024glitchproberadvancingeffectivedetection}
\bibfield{author}{\bibinfo{person}{Zhibo Zhang}, \bibinfo{person}{Wuxia Bai}, \bibinfo{person}{Yuxi Li}, \bibinfo{person}{Mark~Huasong Meng}, \bibinfo{person}{Kailong Wang}, \bibinfo{person}{Ling Shi}, \bibinfo{person}{Li Li}, \bibinfo{person}{Jun Wang}, {and} \bibinfo{person}{Haoyu Wang}.} \bibinfo{year}{2024}\natexlab{}.
\newblock \showarticletitle{GlitchProber: Advancing Effective Detection and Mitigation of Glitch Tokens in Large Language Models}. In \bibinfo{booktitle}{\emph{The 39th IEEE/ACM International Conference on Automated Software Engineering (ASE)}}.
\newblock


\bibitem[Zhou et~al\mbox{.}(2019)]%
        {zhou2019completing}
\bibfield{author}{\bibinfo{person}{Zili Zhou}, \bibinfo{person}{Shaowu Liu}, \bibinfo{person}{Guandong Xu}, {and} \bibinfo{person}{Wu Zhang}.} \bibinfo{year}{2019}\natexlab{}.
\newblock \showarticletitle{On completing sparse knowledge base with transitive relation embedding}. In \bibinfo{booktitle}{\emph{Proceedings of the AAAI Conference on Artificial Intelligence}}, Vol.~\bibinfo{volume}{33}. \bibinfo{pages}{3125--3132}.
\newblock


\bibitem[Zou et~al\mbox{.}(2023)]%
        {zou2023representation}
\bibfield{author}{\bibinfo{person}{Andy Zou}, \bibinfo{person}{Long Phan}, \bibinfo{person}{Sarah Chen}, \bibinfo{person}{James Campbell}, \bibinfo{person}{Phillip Guo}, \bibinfo{person}{Richard Ren}, \bibinfo{person}{Alexander Pan}, \bibinfo{person}{Xuwang Yin}, \bibinfo{person}{Mantas Mazeika}, \bibinfo{person}{Ann-Kathrin Dombrowski}, {et~al\mbox{.}}} \bibinfo{year}{2023}\natexlab{}.
\newblock \showarticletitle{Representation engineering: A top-down approach to ai transparency}.
\newblock \bibinfo{journal}{\emph{arXiv preprint arXiv:2310.01405}} (\bibinfo{year}{2023}).
\newblock


\end{thebibliography}

\end{document}